\documentclass[nolinenumbers]{aastex631}

\usepackage{multirow}

\usepackage{amsmath}
\usepackage[ruled,vlined]{algorithm2e}
\usepackage{rotating}
\usepackage{hyperref}
\usepackage{placeins}
\usepackage{afterpage}
\usepackage{float}
\usepackage{here}
\usepackage{placeins}

\shorttitle{Zorro southern binaries}
\shortauthors{Mendez et al.}

\begin{document}

\title{Southern binaries with the Zorro Speckle Camera @ Gemini-South}

\author[0000-0003-1454-0596]{Rene A. Mendez}
\affiliation{Departamento de Astronomia \\
Universidad de Chile \\
Casilla 36-D, Santiago, Chile}

\author[0000-0002-2084-0782]{Andrei Tokovinin}
\affiliation{Cerro Tololo Inter-American Observatory\\NOIRLab\\Casilla 603, La Serena, Chile}

\author{Edgardo Costa}
\affiliation{Departamento de Astronomia \\
Universidad de Chile \\
Casilla 36-D, Santiago, Chile}

\author[0009-0000-5181-7924]{Maximiliano Dirk}
\affiliation{Centre for Astrophysics Research\\University of Hertfordshire\\ Hatfield, Hertfordshire AL10 9AB, UK}

\begin{abstract}
We present measurements in the context of a survey of southern hemisphere binary and multiple stellar systems observed with the Zorro Speckle dual diffraction-limited optical imaging camera on the 8.1~m Gemini-South telescope carried out between 2019 and 2023. The overall motivation of our survey, as well as some initial results of these observations, are outlined to demonstrate the capabilities - and limitations - of Zorro. We report on the astrometric characterization of the instrument in terms of the precision and accuracy of our measurements and provide details of our custom-made data reduction pipeline. For targets with separations smaller than 0\farcs4, an overall precision of 1\,mas in the radial and tangential directions is obtained, while the uncertainty in position angle is 0\fdg2. Relative astrometry and contrast brightness in the two Zorro filters at 562 and 832 nm are reported for 70 pairs on 64 distinct systems (six are triples). Eleven new binaries are found, mostly of small separations (down to 15\,mas), and large brightness contrast (up to $\Delta m=6$ in the red channel). Our results indicate that the Zorro instrument, when properly calibrated, delivers excellent quality data for visual binary studies of tight and/or faint companions.



\end{abstract}

 
\section{Introduction} \label{sec:preamble}

Vogt-Russell’s theorem \citep{1972A&A....20..105K} predicts that the most fundamental parameter determining the internal structure and evolution of stars of a given chemical composition is their initial mass (see e.g. \citet{2001eaa..bookE1882M, 2013sepp.book.....I}). One of the main relationships reflecting the dependency of the star's properties on mass is the mass-luminosity relation  (MLR), first discovered empirically in the early  20th century, and later explained on  theoretical grounds by  \citet{1924MNRAS..84..308E}. Improving the observational  MLR  is not a simple task,  because it involves not only determining precise distances\footnote{A classical problem in binary-star research, now largely solved by Gaia.},  but also another elusive parameter: mass. To further complicate things, the observational  MLR  has a statistical dispersion which cannot be explained exclusively by observational errors in the luminosity or mass; there seems to be an intrinsic dispersion caused by differences in age and/or chemical composition (see e.g. \citet{2012Ap&SS.341..405G}).\\
 

Currently, the best MLRs for main sequence stars are those of \citet{2010A&ARv..18...67T} and \citet{2016AJ....152..141B}, but (except for one object with [Fe/H]~$< 0.25$ in \citet{2010A&ARv..18...67T}) neither of them include low-metallicity stars. Although there are other studies, using long-base optical interferometry of binary systems, that have begun to address metallicity effects (e.g. \citet{2012ApJ...746..101B, 2012ApJ...757..112B}, \citet{2012ApJ...757...42F}), these have reached only as low as about [Fe/H] = -0.5.

Another pioneering effort to this end, which in fact provided the motivation for the present survey, has been that of \citet{2015AJ....149..151H, 2019AJ....157...56H} who have been using the high-resolution Speckle camera Alopeke mounted on the Gemini-North 8.1~m telescope (GN hereafter) to determine an empirical low-metallicity MLR for the Solar Neighborhood. So far, that work has resolved over twenty systems
that span a range of Iron abundance from [Fe/H] of +0.1 down to -2.0, with spectral types that range from mid-F to mid-K\footnote{A more recent study by \citet{2019ApJ...871...63M} discusses the effects of metallicity on the MLR for later spectral types and masses $M< 0.7$~M$_\odot$.}.

The dependence of total mass on metallicity (for a given luminosity) for the eleven systems with the best data that they had available is shown in Figure~7 of \citet{2019AJ....157...56H}. Alopeke observations on GN for these eleven binaries with a reliable mass determination confirm that the data appears to follow the theoretically expected trend in stellar mass as a function of metallicity. This was a significant step forward, albeit further data are clearly needed to fully constrain the models.

This latter reality prompted us to start in 2019 an observational campaign in the southern hemisphere using the Zorro Speckle camera (a twin of Alopeke) on the Gemini-South 8.1~m telescope (GS hereafter), to increase the sample of low-metallicity objects with well determined orbits and mass sums. Zorro@GS provides a unique opportunity to add an important number of southern systems that will fill in the current metallicity range, and to provide further data on objects that have been scarcely observed at GN. With the full data set, it will be possible to make a relevant contribution to the main-sequence MLR for metal-poor stars.\\

Determining high-quality individual masses is time-consuming, and requires precise astrometric (mass sum) and high-resolution spectroscopic (mass ratios) observations that must span a significant part of the orbital period. Direct determination of individual masses of spectroscopic binaries is possible if the components can be resolved and their angular separation in the sky (denoted by $\rho$ hereafter) can be accurately determined over time. For double-lined systems, the combined spectroscopic/astrometric orbit solution yields individual masses as well as a distance to the system without the need for parallax measurements (the so-called self-consistent “orbital parallaxes”, \citet{2022AJ....163..118A}). An independent distance measure (e.g., from Gaia) is needed for single-lined systems to complete the path to individual masses. With a spectroscopic orbit and parallax in hand, several resolved observations adequately spread out along the orbit can be used to determine the semi-major axis and inclination reliably, and therefore provide the basis for individual masses (\citet{2023AJ....166..172A}).

Given the small space density of local Halo stars, metal-poor binary systems are typically farther away from the sun than solar metallicity stars and, therefore fainter and/or more compact spatially, making them difficult objects for optical interferometry with small (4\,m or less) telescopes. Thanks to their dual-detector design, Zorro@GS and Alopeke@GN have the ability to resolve binary systems even slightly below its natural diffraction limit of ~20~milliarc-seconds (mas). These speckle cameras provide an excellent opportunity to make relatively quick progress on a number of low-metallicity binary systems (given the tight and hence short periods involved), by combining conclusive observations obtained at Gemini with spectroscopic data and lower-precision astrometry from other facilities, already available from the literature.

It is extremely important to precisely calibrate masses and luminosities of metal-poor stars. Typically, the Population II main sequence has been defined by nearby metal-poor stars (e.g. \citet{1997AJ....114..161R}, \citet{1997ApJ...491..749G}), a number of which could be binary systems. If binaries currently included in the Population II main sequence definition are resolved, and individual luminosities can be obtained, these new data will reduce the current scatter, allowing for more stringent constraints on stellar models, as well as better ages and distances to Galactic globular clusters.

In cases where one component has evolved away from the main sequence, age determinations are also possible using, e.g., the method of \citet{2009AJ....138.1354D} to place the components on the H-R diagram. A hint of this is shown in Figures~1 and~8, on Sections~1 and 4, respectively, of \citet{2017AJ....154..187M}, and on Figs. 5 and 6, on Sections 4 and 5 of \citet{2022AJ....163..118A}. Also, the secondary components of metal-poor binaries are especially important in that they will have undergone considerably less change in color and luminosity, and their current observables should thus be close to their zero-age locations in the color-magnitude diagram, and in this way, speckle observations could be directly compared with stellar models \citep{2013ApJ...776...87S}.  Thus, our Zorro@GS survey, together with the Alopeke@GN program, will not only add a significant number of new points to the MLR, but also provide sensitive tests of stellar evolution theory. We expect to be able to investigate the effect of metallicity and age on the MLR for the first time. Also, in general, an increase in the number of well-studied binary stars will also contribute to other astrophysical areas, such as star formation and comprehensive studies of the Solar Neighborhood, which require a knowledge of the multiplicity fraction.\\

Several low-metallicity binary stars targeted by the Alopeke@GN survey turned out to be multiple systems (some pending confirmation of common motion). These objects are quite interesting because the architecture of stellar hierarchical systems results from their formation and early evolution; thus their study helps to understand the formation of stars and planets \citep{2017ApJS..230...15M, 2018ApJ...854...44M}. We have included these compact hierarchical systems in our program, in order to provide further observational constraints on the orbital architecture. In principle, both inner and outer orbits could be determined, and our goal in this respect is to increase the still modest number of multiples where both orbits are known precisely (see, e.g., \citet{2020AJ....160..251T}), enabling the study of relative orbit orientation, mutual resonances, and eccentricities in multiple systems.\\






In this paper, we report on the preliminary results from our Zorro@GS survey and its medium-term prospects. In a forthcoming paper, we will analyze in detail a few of the objects for which enough data permits a precise determination of their orbits, including radial velocities (RV hereafter) from an ongoing effort with Echelle spectrographs to determine RVs for the system's components.

The outline of the paper is as follows: In Section~\ref{sec:sample} we describe the selection of our sample and provide a log of our observing runs. In Section~\ref{sec:redux}, we describe the data reduction steps, including the astrometric calibration and an assessment of the data quality and overall performance of Zorro@GS. In Section~\ref{sec:res} we highlight our main results and provide comments on selected individual objects. Finally, Section~\ref{sec:concl} presents our main conclusions.

\section{Sample selection} \label{sec:sample}

Our initial sample consisted of assorted southern hemisphere targets kindly provided by E. Horch and collaborators, from their study of low-metallicity binaries that lacked orbital coverage or were pending confirmation of their architecture (actually, at least three of the systems targeted by Horch and collaborators were discovered to be trinaries; sustained observations of the new components (if bound) would further shed light to our knowledge of metal-poor stars).

This list quickly evolved in time as observations progressed and some targets were removed from the observing list (e.g., due to non-detection or possibly very long periods), while other promising low-metallicity or multiple-system targets were added from our own survey of southern binaries. Indeed, in 2014, we began a systematic campaign with the HRCam speckle camera at the SOAR 4.1\,m telescope to observe nearby southern binary and multiple stellar systems (both visual and spectroscopic). The science goals of that project have been described in \citet{2017AJ....154..187M}. Many publications have resulted from this effort (for the latest, see e.g.,\citet{2023AJ....166..172A}).

Among the stars observed with HRCam@SOAR, we have identified many tight binary systems that are at the resolution limit of this instrument and, therefore, have large astrometric uncertainties. These are short-period systems ($\sim$10\,yr or less) with very small projected $\rho$, equal to or smaller than the diffraction limit of the SOAR telescope ($\sim$35\,mas at 5500\,Å), in which cases speckle observations with 4\,m facilities are of insufficient precision to determine high-quality orbits. Therefore, in addition to including bona-fide low-metallicity targets in our GS program, we also included some of these tight systems. We have also included some targets from the survey of multiple stellar systems in the southern hemisphere being carried out by A. Tokovinin (see, e.g., \citet{2023AJ....165..220T}) which proved to be too tight for the resolution of HRCam@SOAR.

We must emphasize that, as a result of our selection process, our final sample is very heterogeneous and it should not be considered complete or representative of these systems in any astrophysical sense. From this point of view, this work's main contribution is adding new orbits and mass ratios (when possible) for a variety of tight binaries and multiple systems.

\subsection{Observing strategy and observing runs}\label{sec:runs}


Zorro is a fast, low-noise, dual channel and  dual-plate-scale imager, which in speckle mode provides simultaneous two-color (blue  and  red) diffraction-limited  optical imaging (FWHM~20\,mas at 650\,nm) of objects as faint as $V \sim 17$ mag over a FOV of 6\farcs7. The detector consists of  two $1024 \times 1024$ iXon Ultra 888 back-illuminated  Electron  Multiplying CCDs with 13\,$\mu$m pixels\footnote{Full description of the instrument can be found in: \url{https://www.gemini.edu/instrumentation/alopeke-zorro}}. In speckle mode, four narrow band filters are available\footnote{Narrow band filters are preferred to increase the speckle contrast.}, two of which can be used simultaneously in the blue and red "arms" of the instrument. For our survey we selected the Edmund Optics filter \#562 in the blue side (562\,nm central wavelength, 54\,nm wide), and filter \#832 (832\,nm central wavelength, 40\,nm wide) on the red side.\\

In general, we followed the standard recommendations given on the Gemini web page for the instrument, but added two sensitive observing constraints. We requested that airmass be kept as small as possible and that the observations be limited to hour angles within $\pm$1.5 hours from the meridian to minimize the effects of atmospheric dispersion (AD), which was especially noticeable in the blue filter (see Section~\ref{sec:binproc}). Unfortunately, due to the queue nature of the observations, combined with the declination of some targets, this rule was not always possible to strictly enforce.

In our list of targets we included well-observed, relatively wide, binaries for calibration purposes (further details are given below), which were typically observed at a rate of at least two systems per run. We also assigned a PSF reference star to each target; these are bona-fide single stars closer than about 3\degr ~to the target used for modelling the power spectrum of an isolated star. While the Gemini Zorro web pages recommend selecting bright objects ($5 < V < 6$ mag), after our first few runs we found that such bright objects did not provide a good match to the binary fit (see Section~{\ref{sec:binproc}), so we opted instead for slightly fainter objects, with $V \ge 6$ mag, selected from the Bright Star Catalog (which is limited to $V \le 6.5$).\\

Between 2019 and 2023, we have been awarded more than 74 hours of mostly highest-priority (Band~1) time with Zorro@GS through highly rated proposals by the time allocation committee, but, due to the COVID pandemic, a fraction of this time ($\sim$30\%) could not be executed, significantly delaying our original observing plan. Very high-quality observations have been secured, however, for several tight binaries, and we do have a few new orbital points for them, but in some cases, it has not been possible to obtain the number of epochs needed to derive reliable visual orbits.

In Table~\ref{tab:obslog}, we present the details of our eleven observing runs, including the number of hours allocated, the completion rate, the epochs involved, the total number of usable frames, and the number of individual targets observed.


\begin{table}[ht]
\caption{Observing log from the Gemini Observatory archive: Gemini-Zorro program IDs and dates.} \label{tab:obslog}
\begin{center}
\begin{tabular}{ccccc}
\hline
Gemini Program ID & Allocated time (hrs)/ & Dates & Number of frames\tablenotemark{a} & Number of targets\tablenotemark{b}\\
 & Completion rate (\%) &  & \\
\hline
GS-2019A-Q-110 & 8.5 / 100 & 2019-05-19 to 2019-07-19 & 315 & 38 \\
GS-2019A-Q-311 & 2.9 / 66 & 2019-05-21 to 2019-06-20 & 34 & 7 \\
GS-2019B-Q-116 & 8.9 / 78 & 2019-09-12 to 2019-01-15 & 288 & 41 \\
GS-2019B-Q-223 & 2.3 / 26 & 2019-09-13 to 2019-10-10 & 22 & 4 \\
GS-2020A-Q-116 & 8.0 / 58 & 2020-03-15 to 2020-03-12 & 256 & 34 \\
GS-2020B-Q-142 & 9.2 / 99 & 2020-10-29 to 2021-07-21 & 333 & 42 \\
GS-2021A-Q-141\tablenotemark{c} & 7.7 / 87 & 2021-01-14 to 2021-07-19 & 160 & 44 \\
GS-2021B-Q-145 & 5.7 / 99 & 2021-09-18 to 2021-12-22 & 116 & 38 \\
GS-2022A-Q-150 & 7.9 / 95 & 2022-03-15 to 2022-05-18 & 116 & 38 \\
GS-2022B-Q-143 & 6.8 / 93 & 2022-10-07 to 2023-01-09 & 191 & 58 \\
GS-2023A-Q-142 & 7.3 / 86 & 2023-03-05 to 2023-07-05 & 172 & 47 \\
\hline
\end{tabular}
\end{center}
\tablenotetext{a}{Usable frames ($256 \times 256$ pix$^2$).}
\tablenotetext{b}{Both filters. Includes astrometric calibration binaries (typically two per observing run), and PSF reconstruction bona-fide single stars (typically one per object, either target or calibration binary).}
\tablenotetext{c}{This run was split into two. In the second one the camera on the blue channel was out of focus, and no data was acquired with it. For calibration purposes we used the same values as those derived from the first run of this semester.}



\end{table}

\section{Data reduction, calibration, and precision assessment} \label{sec:redux}

\subsection{Introduction and data}
\label{sec:intro}

In this section, we describe an adaptation of the HRCam@SOAR\footnote{For information regarding this instrument, see \url{https://www.ctio.noirlab.edu/~atokovin/speckle/index.html}} speckle pipeline made to process the data secured with the Zorro@GS. This pipeline is described in \citet{2010AJ....139..743T} and \cite{2018PASP..130c5002T}. We note that the SOAR pipeline has also been recently used to process the data from NESSI, a similar 2-channel speckle instrument at the WIYN telescope \citep{2019AJ....158..167T}.
  
The pipeline was adapted by one of the authors (AT) using as test bench  the first-epoch data from 2019A, which serve here as example. Initially, the Zorro data (being a visiting instrument) was not ingested regularly to the Gemini Observatory Archive; therefore, these data were instead kindly provided by the Zorro instrument scientist at GS, Dr. R.~Salinas, as tar-files, one per night. There are 8 nights from 2019-05-19 to 2019-07-19 (see Table~\ref{sec:runs}). Each tar file contains FITS data cubes compressed with {\tt bzip2}. A standard cube has a format of 256~pix $\times$ 256~pix $\times$ 1000~frames (comprising one data cube), with a file size of 131~GB.
The compressed data cubes have only a slightly smaller size (between 80 to 100 Gb), so the compression does not save much disk space but adds time for a decompression. The files are named like {\tt S20190519Z0318b.fits} and {\tt S20190519Z0318r.fits} for the blue and red channels, respectively, and include the date and a sequential number. The naming is important for the data organization, and the code has been properly adapted to it.

\subsection{Pipeline presentation}\label{sec:pipe}

\begin{figure}[ht]
\epsscale{1.8}
\centerline{
\includegraphics[width=10cm]{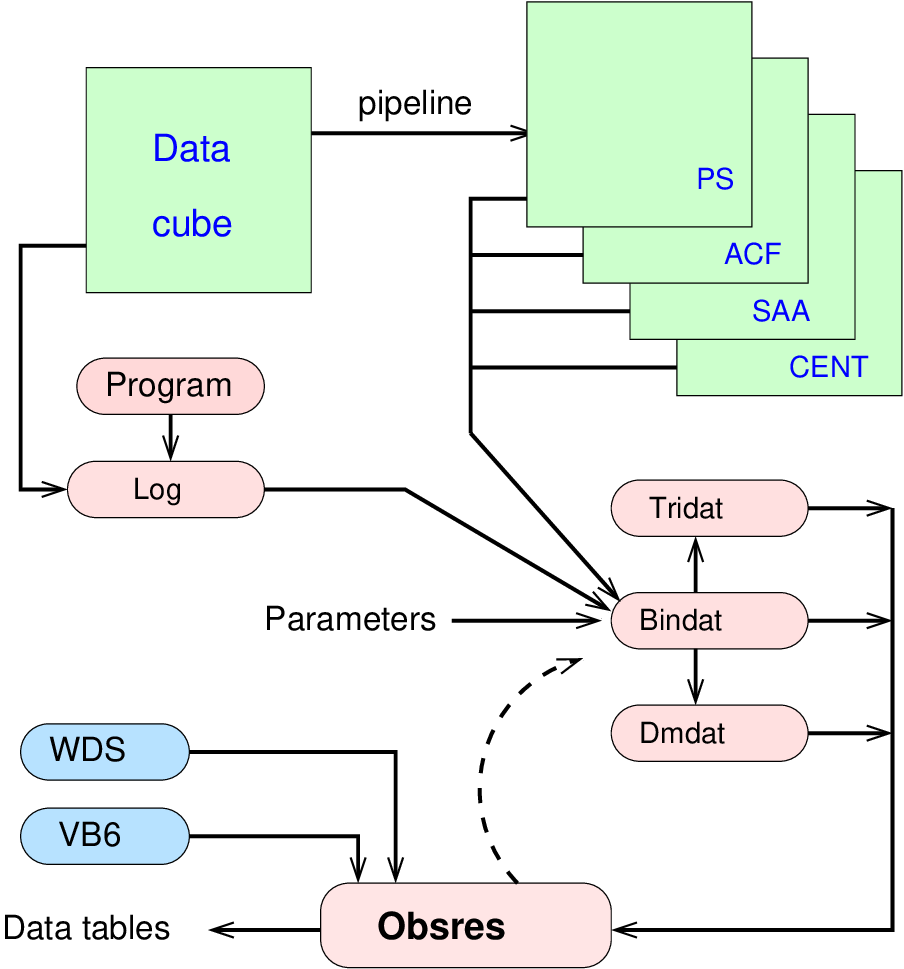}
}
\caption{Data flow diagram of the Zorro Speckle data reduction process, which has been adapted from the SOAR/HRCam pipeline \citep{2018PASP..130c5002T}.}
\label{fig:dataflow}
\end{figure}

The pipeline is written in IDL\textregistered~(version 7.1 or  higher\footnote{IDL is a product by NV5 Geospatial software, see \url{https://www.nv5geospatialsoftware.com/}}), and requires the ASTRO IDL library\footnote{Maintained by NASA's Goddard Space Flight center, and which can be downloaded from \url{https://asd.gsfc.nasa.gov/archive/idlastro/}}. Our pipeline has been developed for personal use and is not meant to be a commercial software product, which means that a potential user must understand the algorithms and be able to fix problems. In the pipeline, the work is organized by means of arrays of IDL structures, saved on disk (Figure~\ref{fig:dataflow}). The {\tt log} and {\tt bindat} arrays have one element per data  cube and contain all essential meta-information, as well as the results of binary-star processing. In  contrast, the {\tt obsres} array (results) has one element per measurement, normally from averaging several data cubes (see further details below). Another important item is the parameter file that specifies path to the data and results, pixel scale, orientation, etc. Each observing run has its own parameter file, and its results (structures) are saved in a separate directory. A parameter file for one observing run,  {\tt z19a.par}, is reproduced below:

\begin{small}
\begin{verbatim}
{param,
 tel: 'GemS', 		; telescope name
 D:   8.1,              ; aperture diam., m
 pixel: 0.00991, 	; nominal pixel scale, arcsec, blue; 0.1093 red, from Zorro/Gemini web page
 long: -70.73669D0,  ; longitude, degrees, Gemini site
 lat:  -30.24075D0,  ; latitude,  degrees, Gemini site
 altitude: 2722., 	  ; altitude a.s.l. for refraction, Gemini page
 dirdat: '/19a/Raw/', ; image files
 dirps:  '/19a/Reduced/ps/',  ; power spectra directory
 diracf: '/19a/Reduced/acf/', ; ACFs directory
 dirav:  '/19a/Reduced/av/',  ; average directory
 filters: ['562','716','832'], ; filter names
 lambda: [562.3,716.0,832.0],  ; central wavelength for each filter, nm
 dlambda: [43.6,51.5,40.4],    ; FWHM, nm
 paoffset: +0.42,  ; add to get true PA                       **Always start with this value at 0.0!
 scale:    0.9606, ; multiply by scale to get true separation **Always start with this value at 1.0!
 redbluescale: 0.9719,     ; pixel(red)/pixel(blue)           **Always start with this value at 1.0!
 redbluetheta: 0.439,      ; theta(red)-theta(blue)           **Always start with this value at 0.0!
 log:      '/19a/Results/log.idl',      ; log-file
 bindat:   '/19a/Results/bindat.idl',   ; bindat structure
 tridat:   '/19a/Results/tridat.idl',   ; tridat structure
 wdsid:    '/19a/Results/wdsid.txt',    ; created by avres2, edited
 avres:    '/19a/Results/avres.idl',    ; averaged results
 obsres:   '/19a/Results/obsres.idl'}   ; final results
\end{verbatim}
\end{small}

\subsection{Data processing} \label{sec:proc}

\subsubsection{Calculation of the power spectra}

The SOAR routine {\tt getpowerixon.pro}, adapted to  NESSI, was used for Zorro without changes. To subtract the bias, we compute the median signal in 10$\times$10 pixel boxes in the lower-left and top-right corners and take the smallest of the two numbers as the (fixed) bias value. 


The main calculation proceeds in the same way as done at SOAR. During the first pass through the cube, the bias is subtracted from each frame,
centroids and other parameters are computed and saved, and the bias-subtracted cube is created. On the second pass, the power spectrum is accumulated using this cube. Also, the centered and "shift and add" (SAA hereafter) images (useful for quadrant disambiguation) are computed. The individual image parameters were used to reject poor frames (e.g. target too close to the border). Of the full 1024$\times$1024~pix detector we used only a small region of interest (ROI) of 256$\times$256~pix, so the fraction of rejected images is substantial for some cubes.


Also (especially in the earlier runs)  a few data cubes were recorded with the full FOV on each night. These huge cubes (16 times normal) are rejected by the power calculation and not processed any further (these are considered unusable frames, and are excluded from Table~\ref{tab:obslog}).\\

Due to its optical layout, the two Zorro detectors have different orientations on the sky. To avoid potential confusion in further data reduction, we flip the red channel images along the horizontal axis, leaving the blue channel unchanged. As a result, all images have the same orientation: North to the right, East down. The resulting 2D images are named using the standard prefixes ps-, av-, cent-, saa-, and can be automatically retrieved during data processing from the names stored in the {\tt log} or {\tt bindat} structures. The results of {\tt getpowerixon} are placed in the directories specified by the parameter file (all nights for a given observing run are put together).

\subsubsection{The log structure and the parameter file}

The log structure, identical to that of SOAR, is created and filled by several procedures in {\tt getlog.pro}. The first procedure, {\tt getlog}, reads all FITS files in the chosen directory, extracts relevant parameters from the headers, and creates the log structure, one element per data cube. The structure is saved in the log file specified by the parameters, or appended to it if the keyword {\tt /append} is given. The log is created by calling the program for each night and appending successive nights of the same run. The filter names in the log (strings) equals the wavelength in~nm, e.g. '562'. The procedure  {\tt getzen} computes the zenith distance and  parallactic angle, using star coordinates and the geographical coordinates of the site in the parameter file.
The routine {\tt listlog} in {\tt getlog.pro} creates the text  {\tt listlog.txt}, to be consulted during binary processing (e.g. to find a proper PSF reference star for a given binary). For example, for 2019a, the log contains 456 lines, including the full-frame cubes not processed here.

The code {\tt getacf.pro} computes all ACFs from the power spectra. It operates on the log structure and adds to it important parameters such as models of the power spectrum (used in the binary-star processing), detection limits, and parameters of the AD. At this point, the log structure is complete. It is copied under the name {\tt bindat} and used in the binary-star processing, which is the next step.

\subsubsection{Binary star processing} \label{sec:binproc}

\begin{figure}[ht]
\epsscale{1.8}
\centering
\includegraphics[clip, trim=60 350 150 250, width=0.8\textwidth]{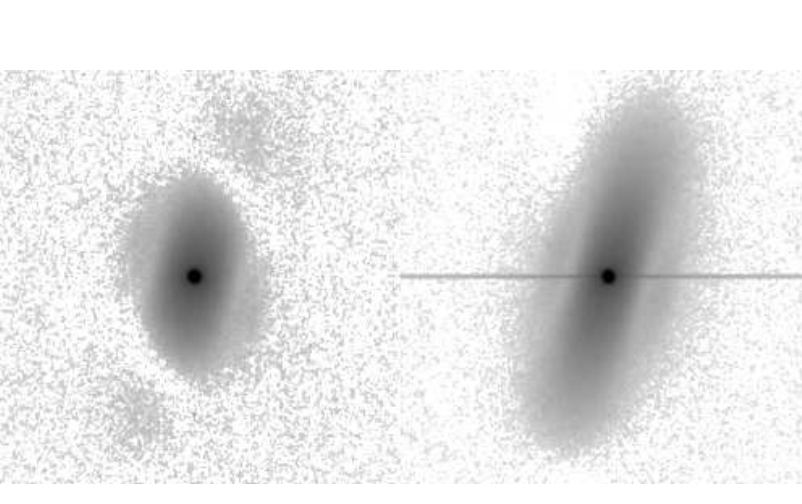}\\
\includegraphics[clip, trim=60 350 150 250, width=0.8\textwidth]{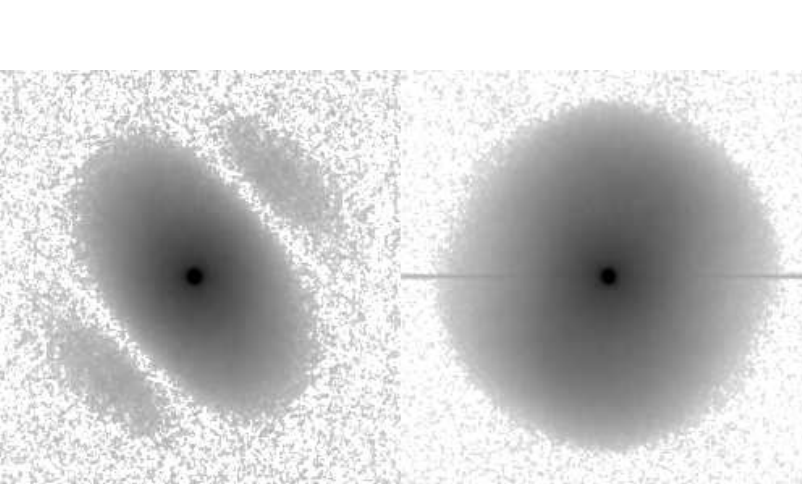}
\caption{Example of power spectra for the data cube {\tt S20190519Z18}, displayed with a negative logarithmic stretch. Top: blue channel, bottom: red channel. To the right of each spectrum, the spectrum of the associated PSF reference star is shown. The object is the 21\,mas separation binary HIP~89000 observed at a zenith distance of 35$\degr$. The horizontal line in the reference spectra is an artifact caused by the image truncation owing to imperfect centering of the star in the ROI. 
\label{fig:spectra} }
\end{figure}


The SOAR procedure for binary-star fitting {\tt bin8.pro} and its GUI interface  {\tt xb.pro} were slightly adapted. The default reference spectrum calculated by azimuthal average \citep{2010AJ....139..743T} includes modeling of the AD. However, the Zorro filters have a rectangular passband, hence their transfer function is modeled by the sinc$^2$ function rather than by a Gaussian. Figure~\ref{fig:spectra} shows the power spectra of a tight binary (HIP 89000) in the blue and red channels. The power spectrum of the associated nearby PSF reference star is shown on the right side. Unfortunately, we found that the resolution in the blue channel is severely degraded by AD in one direction\footnote{Unfortunately, GS lacks an AD corrector, which would greatly alleviate this problem, as has been demonstrated in the case of HRCam@SOAR \citep{2016SPIE.9908E..3BT}.}. If the 21\,mas separation binary were oriented parallel to the AD, it would probably remain unresolved in the blue channel. The situation is much better in the red channel with the 832\,nm filter. In fact, the red channel resolution {\it always} supersedes the blue channel resolution, despite the longer wavelength. The formal astrometric and photometric accuracy (from the binary fits) are also worse in the blue channel. Note also that the blue image is under-sampled by the Zorro pixels.

\begin{figure}[ht]
\centerline{
\includegraphics[clip, trim=60 350 150 250, width=0.8\textwidth]{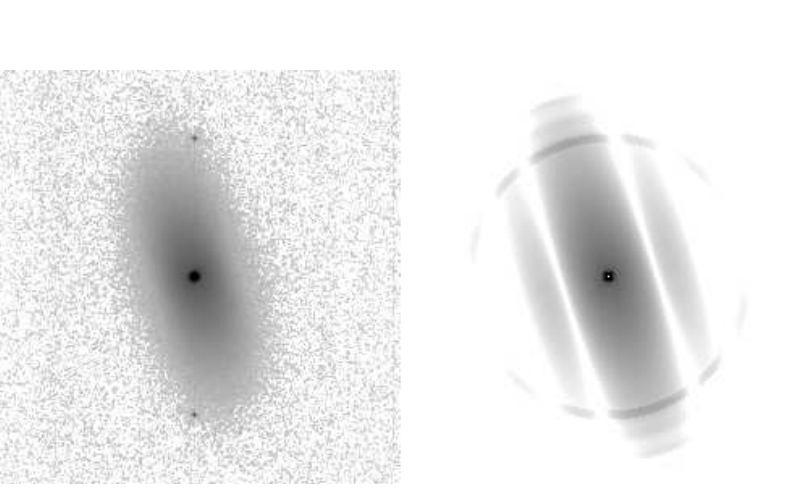} 
}
\caption{Left panel: Power spectrum in the blue channel of the cube {\tt S20190717Z0990b}. Black dots are the consequence of a periodic noise. In this case, the reference spectrum on the right is the RAD model. Its similarity to the object spectrum illustrates the correctness of the AD model.
\label{fig:stripes} 
}
\end{figure}

The use of a PSF reference star takes the AD and other instrumental effects into account, at least to first order. When a PSF reference star is not available, the reference spectrum is constructed from the object itself, by angular averaging, taking into account the AD (we call this the "RAD" model). Figure~\ref{fig:stripes} shows an example of the blue channel power spectrum with a RAD reference. The AD is modeled using the calculated parallactic angle, zenith distance, and filter parameters. In some cases, the binary star fit using a RAD reference gives smaller residuals than the fit using a reference star, and then RAD is preferred. 

Several data cubes taken in 2019 July, including the one shown in Figure~\ref{fig:stripes}, are affected by periodic noise in the form of horizontal stripes. It is manifested in the power spectrum by two strong peaks above and below the center (see the left panel). The peaks create a "ring" in the reference spectrum (right panel) and stripes in the ACF. Such peaks can be easily masked, but this has not been implemented so far in our pipeline. None of the iXon CCDs used so far have had similar defects; and in Zorro, it is also intermittent. The red channel is affected by the stripes at the same time, but to a smaller degree.

The quality of the binary fit is characterized by the reduced $\chi^2/N$. This metric rarely reaches small values, close to one or two. The residuals are typically dominated by the large-scale mismatch between data and model. When the  $\chi^2/N$ with a real PSF reference star exceeds the $\chi^2/N$  with the RAD reference, the latter is preferred. During data processing, the reference star is identified by the cube number only, given by the 4 digits of the file name following 'Z', e.g. '0284'. The function {\tt reference()} in {\tt xb.pro} finds the name of the PSF reference image, given the name of the target file. It automatically selects the correct channel using the 'b' and 'r' letters in the file name.

Several data cubes are usually taken for each binary, producing e.g. a 'b-r-b-r-b-r' sequence in the log file. To automate the processing, we fit the first data cube and use the command {\tt clone} for the following cubes. It copies the results (including the reference name) from the previous {\tt bindat} record to be used as an initial approximation for the fit. Considering that the blue and red channels are different, we typically used {\tt clone, s=-2} to copy from 2 cubes before, i.e. in the same channel. The fitting of triple stars is adapted in a similar way and does not present any additional problems. The final reduced data are stored in the sub-directory {\tt RUN/19a/} in files {\tt bindat.idl} and {\tt tridat.idl}.


The IDL procedure {\tt av} in {\tt avres2.pro} averages the results from similar data cubes (per epoch and filter), producing one record for each subsystem in each filter. Another procedure {\tt wdsid} creates the 'dictionary' file {\tt wdsid.txt} to translate object names to standard WDS discoverer designation codes\footnote{See the on-line WDS catalog at \url{http://www.astro.gsu.edu/wds/}} (when they are not available, the original names are retained). Then a group of procedures in {\tt obsres.pro} creates the {\tt obsres} structure with final results, which is used in further analysis. The data browser {\tt xd.pro} allows to view and, if necessary, re-process individual data (e.g., for quadrant adjustments).

\subsection{Calibration} \label{sec:cal}

\subsubsection{Matching red and blue channels}\label{sec:matchrb}

\begin{figure}[ht]
\centerline{
\includegraphics[width=8.5cm]{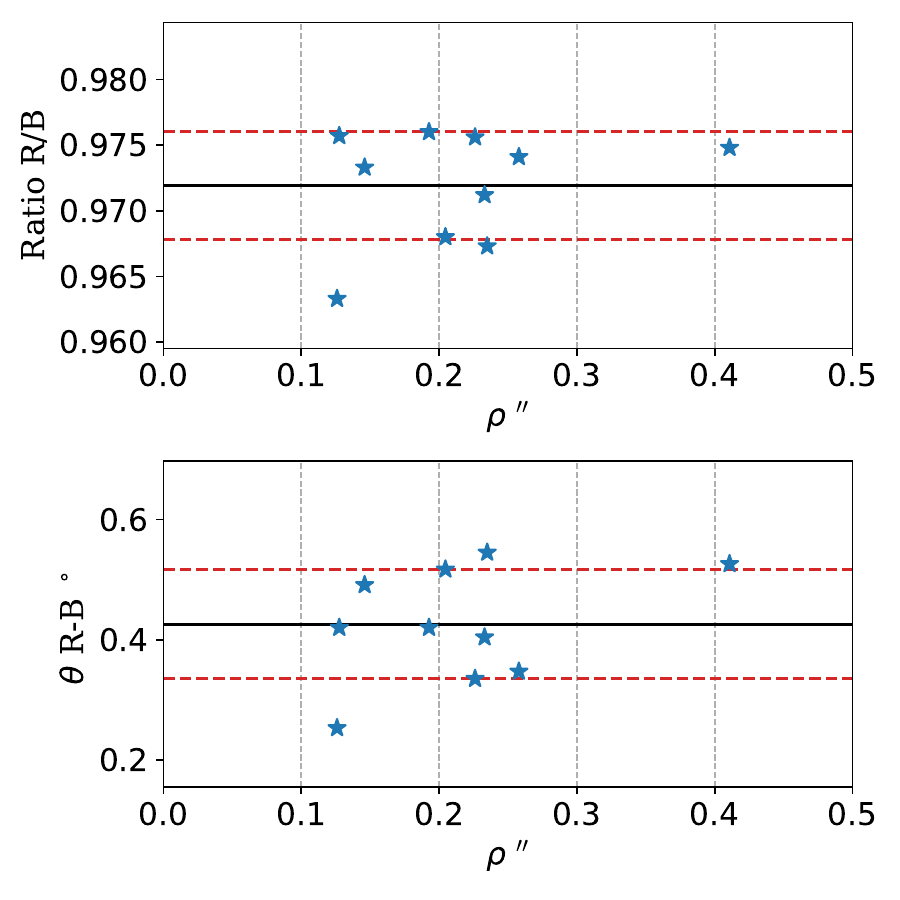} 
}
\caption{Comparison of measurements between the blue and red channels for 10 binary stars observed in the 2019A run. Top panel: ratio of red/blue separations; lower panel: difference in position angle. This comparison is used to correct the red channel to  the same scale and orientation as that of the blue channel. The solid line shows the mean, while the dashed lines indicate the $\pm 1 \sigma$ range.
\label{fig:bluered} 
}
\end{figure}

The detectors of the blue and red channels have slightly different orientation and pixel scales. The nominal values of the pixel scales are 9.91 and 10.93\,mas for the blue and red cameras, respectively, but for the data processing we used a common pixel scale of 9.91 mas for both channels (see the parameter file in Section~\ref{sec:pipe}). The code {\tt redblue.pro} selects from {\tt  obsres} binaries  with $\rho$ between 0\farcs05  and 1\farcs2 and $\Delta m_b < 5.5$~mag, measured in both channels. In the 2019A run there are ten such pairs. Taking the blue channel as a reference, the ratio of scales is plotted in Figure~\ref{fig:bluered}. The mean $\rho$ ratio red/blue is 0.9719 (with an rms over the ten frames of 0.0044), not the 1.1029 ratio inferred from the nominal pixel scales. The mean difference in the position angle ($\theta$ hereafter), separation-weighted, is 0\fdg439  with an rms scatter of 0\fdg095. The procedure {\tt redcorrect} in {\tt redblue.pro} corrects the data in the red channel to match the blue one. After this correction is applied, the pipeline's measurement errors roughly match the channel's difference if an "instrumental" error offset of 0.4\,mas is added in quadrature. This agreement between the channels is similar to the one found for NESSI, after correcting the inter-channel systematics \citep{2019AJ....158..167T}. The typical number of red and blue pairs in all our observing runs varied between 7 and 17, and the precision of the scale ratio was between 0.12\% for the best correction, up to 0.67\% for the worst correction, while the \(\theta\) uncertainty correction remained below 0\fdg5 for all runs.

\subsubsection{Absolute Calibration}\label{sec:abscal}

For the purpose of absolute pixel scale and sky orientation calibration, binaries with $\rho$ between 0\farcs5 and 1\farcs2 and moderate $\Delta m$ have been selected as "astrometric calibrators". Their motion is typically slow and it can be accurately modeled if good modern data are available. The current standard Zorro calibration plan does not contain such binaries, therefore in each observing run we have included them in our target list. These calibrators have been inherited from the SOAR program. We note that the accuracy of the models has nothing to do with the accuracy of their corresponding orbits that may remain poorly constrained for some time. The SOAR calibrator models have been updated in \citet{2022AJ....164...58T} and references, as a whole, to the relative positions measured by Gaia DR3. If, in the future, more accurate models of the calibrators become available, the observations can be corrected a-posteriori.


In addition to our astrometric calibrators, the 2019A data also includes several binaries with well-known orbits (including two of grade~1), probably added as part of the Zorro standard calibration plan. These latter binaries are mostly tight, $<\,$0\farcs2, and the accuracy of their orbits is always worse than the intrinsic data accuracy (additionally, observations of the two grade-1 binaries at SOAR indicate that both orbits require minor corrections). Moreover, the accuracy (weight) of calibration in scale and orientation is proportional to the binary $\rho$, so binaries with larger $\rho$ should be preferred. Note also that relying on orbits or data from smaller telescopes is a poor calibration strategy.

\begin{figure}[ht]
\centerline{
\includegraphics[width=8.5cm]{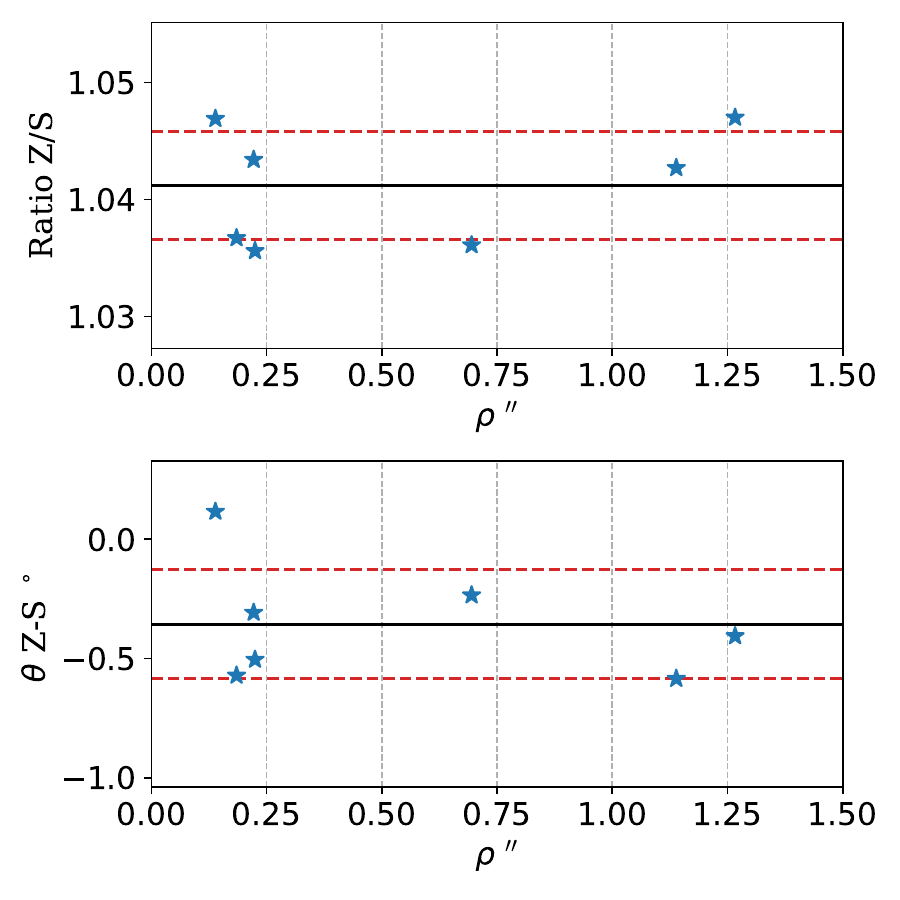}
}
\caption{Calibration of Zorro astrometry relative to SOAR. The top panel shows the scale ratio, while the lower panel shows the difference in $\theta$ for the calibration binaries included in the 2019A run. The solid line shows the mean, while the dashed lines indicate the $\pm 1 \sigma$ range. \label{fig:soar} 
}
\end{figure}

Lacking "perfect" absolute calibrators, we can only compare data from Zorro with those from SOAR. Excluding triples, there are seven systems in common observed at SOAR in 2019A and before. These include our own astrometric calibrators and the tighter calibration binaries provided by the observatory mentioned in the previous paragraph. We first interpolated the SOAR positions linearly to the epoch of the Zorro observations and then compared with Zorro (red and blue channels of Zorro are now in good agreement and are averaged, see previous sub-section). The plots in Figure~\ref{fig:soar} show the comparison between the Zorro and SOAR measurements. The mean $\rho$ ratio Z/S is 1.041 (rms 0.005) and the separation-weighted difference in \(\theta\) is Z$-$S=$-$0\fdg42 (rms 0\fdg25). These corrections were introduced into {\tt obsres} manually (see the parameter file in Section~\ref{sec:pipe}).
The pixel scale of the blue and red Zorro channels resulting from this calibration for the 2019A dataset turned out to be 9.520 and 9.252\,mas, respectively.

The same procedures outlined above were applied to all successive observing runs for which we had always at least two of these astrometric calibrators. The uncertainty of the calibration varied bewteen 0.02\% and 0.50\% in scale ratio, while the \(\theta\) uncertainty correction remained below 0\fdg25.

%
%
%
%
\begin{longrotatetable}
\begin{deluxetable}{cccccccccccccccp{3cm}} 
\tablecaption{Resolved systems.\label{tab:double}}
\tablecolumns{16}
\tabletypesize{\tiny}
\tablehead{\colhead{WDS} &
\colhead{DD} &
\colhead{HIP} &
\colhead{Epoch} &
\colhead{Filt} &
\colhead{Ncub} &
\colhead{\(\theta\) } &
\colhead{$\rho\times \sigma_{\theta}$} &
\colhead{\(\rho\)} &
\colhead{$\sigma_{\rho}$} &
\colhead{$\Delta$m} &
\colhead{Code} &
\colhead{$\Delta\theta$} &
\colhead{$\Delta\rho$} &
\colhead{Ref} &
\colhead{Comments} \\
\colhead{} &
\colhead{} &
\colhead{} &
\colhead{+2000.0~yr} &
\colhead{} &
\colhead{} &
\colhead{\degr} &
\colhead{mas} &
\colhead{\arcsec} &
\colhead{mas} &
\colhead{mag} &
\colhead{} &
\colhead{\degr} &
\colhead{\arcsec} &
\colhead{} &
\colhead{}
}
\startdata
01011+1622 & DSG 9               &  4754       & 20.8246 & 832  & 5 &   35.9  & 10.3 &  1.3461 & 10.5 &    7.5 &   &  &  &  & [Fe/H] = $-$1.72, large proper$-$motion. \\
\hline
01059$-$2529 & HD 6516             &  5146       & 19.5439 & 832  & 3 &  158.1  &  5.0 &  0.1821 &  4.6 &    5.3 &   &  &  &  &  \\
\hline
01205$-$1957 & TOK 203             &  6273       & 21.7198 & 562  & 2 &  127.0  &  2.0 &  0.2707 &  1.7 &    6.1 &    &   17.7 &  0.057 & Tok2019c & \\
           &                     &             & 21.7198 & 832  & 2 &  127.2  &  0.1 &  0.2726 &  0.1 &    3.7 &  q &   18.0 &  0.059 & Tok2019c & \\
\hline
01262$-$6751 & DON 17              &  6703       & 20.8246 & 562  & 5 &  291.0  &  0.2 &  1.4060 &  0.2 &    1.5 &  q &    0.0 & $-$0.008 & Izm2019  &  Exclude, corrupted measurements. \\
           &                     &             & 20.8246 & 832  & 5 &  291.1  &  0.1 &  1.4087 &  0.3 &    1.2 &  q &    0.1 & $-$0.005 & Izm2019  &  Outside/edge FOV! Corrupted ACF. \\
           &                     &             & 21.7200 & 562  & 2 &  248.0  &  0.5 &  1.1937 &  0.1 &    1.8 &  q &  $-$43.2 & $-$0.231 & Izm2019  & \\
           &                     &             & 21.7200 & 832  & 2 &  249.5  &  0.2 &  1.2637 &  0.0 &    1.3 &  q &  $-$41.7 & $-$0.161 & Izm2019  & \\
\enddata 
\end{deluxetable}
\end{longrotatetable}

Since some of our astrometric calibrators were observed over several epochs, we can compare the results of our calibrations over time, which are done independently for each observing run based on the epoch-specific ephemeris. This is shown, as an example, in Figure~\ref{fig:astrombin} for HIP~8998 (three distinct epochs) and HIP~46454 (four epochs), with two measurements per epoch (blue and red, usually indistinguishable at the scale of the plots). The actual data is given in Table~\ref{tab:double}. As seen from Table~\ref{tab:astromcomp}, the calibration renders our measurements consistent with those from the SOAR program. No systematic offsets are evident, and a similar scatter is seen in the zoom inlets of Figure~\ref{fig:astrombin}, which shows that our calibrations have been applied consistently throughout the different epochs. However, given the small number of comparison points, it is difficult to assess the internal precision of the Zorro data from this comparison; this is done in more detail in the following section.

\begin{figure}[ht]
\centerline{

\includegraphics[width=8.5cm]{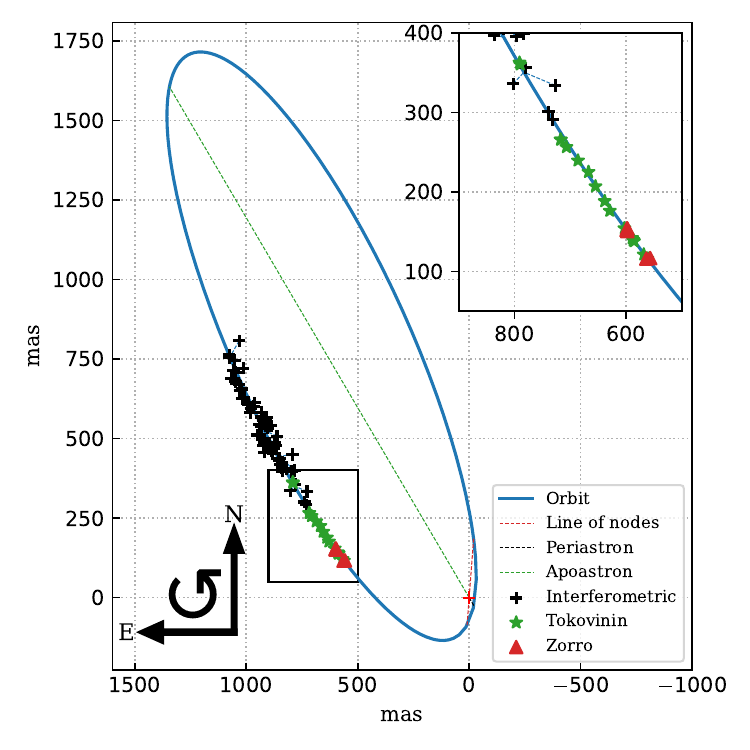} 
\includegraphics[width=8.5cm]{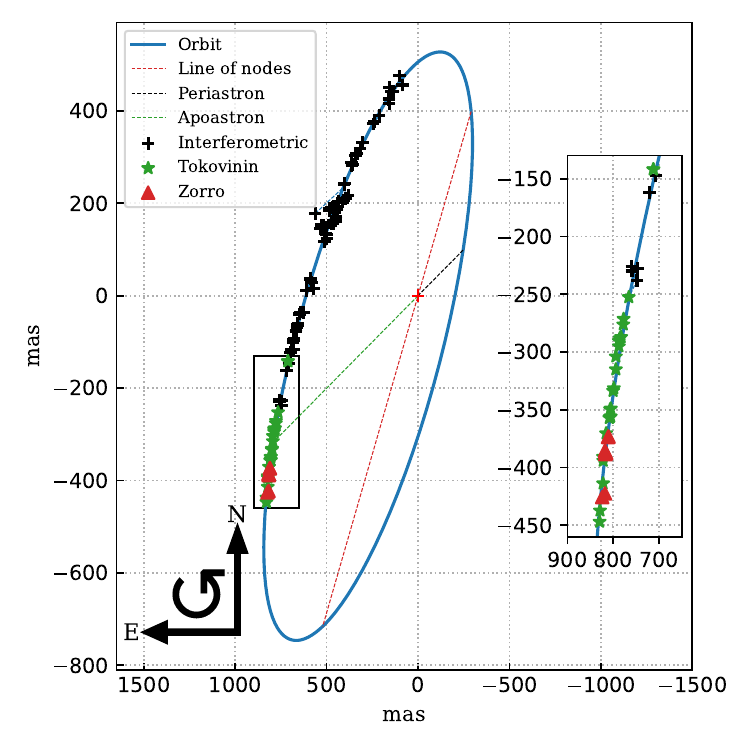} 
}
\caption{Example results for two astrometric calibrators: HIP 8998 (left panel, three epochs) and HIP 46454 (right panel, four epochs). In both cases blue and red measurements could be secured on each epoch (usually indistinguishable at the scale of the plots). The orbits shown are preliminary and are bound to be modified as more data are acquired. Indeed, these two orbits are currently listed as grade 2 in ORB6, with estimated periods of 167\,yr and 118\,yr respectively.}
\label{fig:astrombin} 

\end{figure}

\begin{table}[ht]
\caption{Residuals for SOAR (first line) and ZORRO (second line) for the two astrometric calibrators (HIP 8998 and HIP 46454) whose tentative orbits are shown in Figure~\ref{fig:astrombin}.}
\begin{center}
\begin{tabular}{ccccccc}
\hline\hline
Binary & $[O-C]_{\rho}$ & $\sigma_{\rho}$ & $[O-C]_{\theta}$ & $\sigma_{\theta}$ & Epoch range & Npoints\\
& mas & mas & \degr & \degr & yrs & \\
\hline
8998 & 3.1 & 2.8 & +0.02 & 0.21 & 2008.77 - 2023.00 & 17 \\
    & -2.2 & 2.0 & +0.03 & 0.12 & 2020.91 - 2023.02 & 6 \\
\hline
46454 &  3.2 & 2.0 & -0.02 & 0.17  & 2009.27 - 2024.15 & 25 \\
      & -2.3 & 2.8 & -0.02 & 0.15 & 2020.20 - 2023.01 & 8 \\
\hline\hline
\end{tabular}
\end{center}
\label{tab:astromcomp}
\end{table}

\subsection{Results}
\label{sec:res}

The final results for a particular run are stored in the  {\tt obsres} structure. They can be exported in different formats. The standard export creates the files {\tt  double.txt} and {\tt  single.txt} for resolved and unresolved sources, respectively. For example, in 2019A there is one resolved triple HIP~63377 with $\rho$ of 29\,mas and 0\farcs46 between Aa,Ab and Aa,Ac, respectively (see Table~\ref{tab:double}). Its outer pair is known as TOK~722. A new faint companion to the PSF reference star HIP~82216 (HD 151526) is detected at 0\farcs72. The bright pair STF~1998 (WDS J16044$-$1122) was observed only once and only in the blue channel; its orbit probably needs a correction. We note that in the earlier epochs (including 2019A) an excessive number of data cubes (6 or even 7) were acquired for some targets. We found that these over-abundant data did not improve the quality of the final result but did cost observing time. After the initial runs, we started acquiring typically only two data cubes per target on a regular basis.

\subsection{Data Quality: Internal Precision Assessment} \label{sec:quality}

In this subsection we provide a look at the quality of the data, based mostly on repeated measurements, which gives us an idea of the internal precision of our relative photometry and astrometry. Because all our targets are observed simultaneously in two filters in every epoch, this allows us to estimate the repeatability of our measurements of $\rho$ and $\theta$ (which, in principle, should be independent of the filter used). Because we have observations at different epochs for several systems, we can also assess the precision of the magnitude difference between the components --- which should be independent of the epoch of observation --- unless, of course, one of the binary components is variable.

\begin{figure}[ht!]
\epsscale{1.1}
\plotone{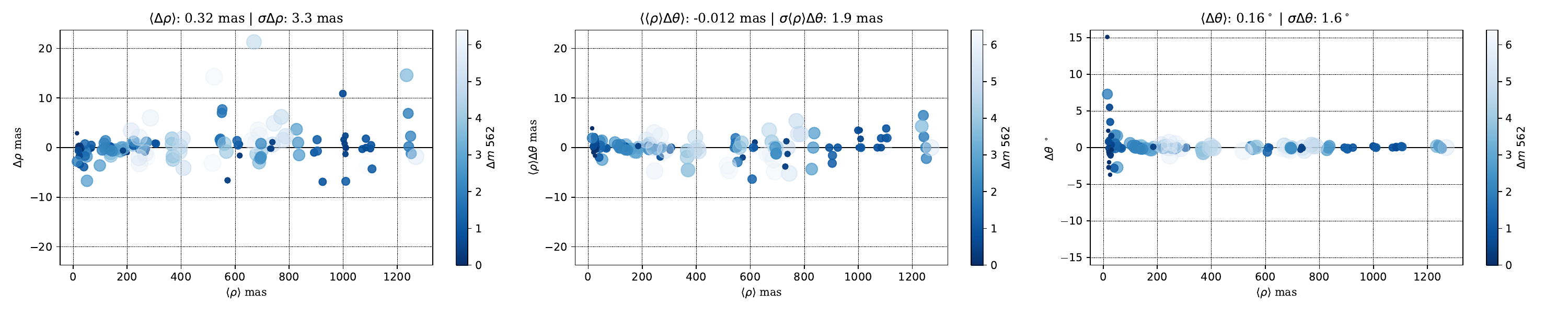}
\plotone{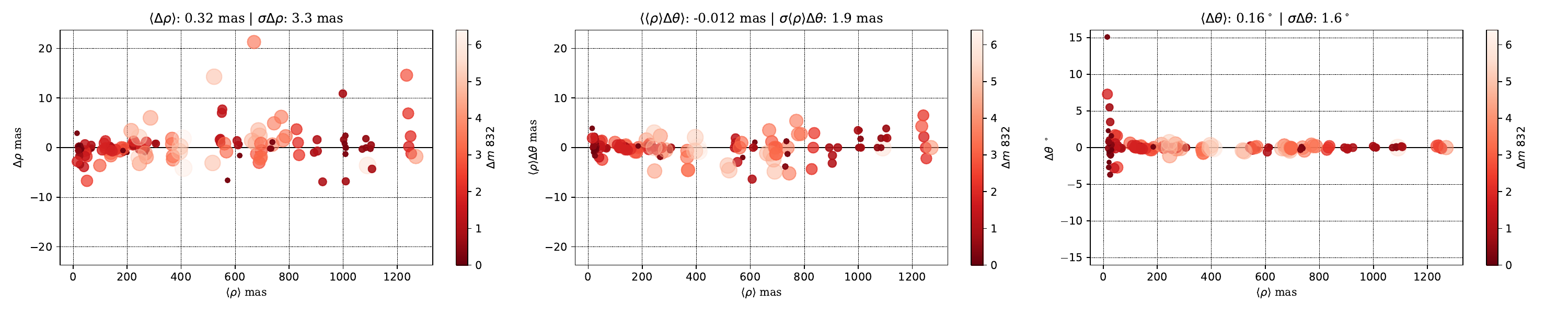}
\plotone{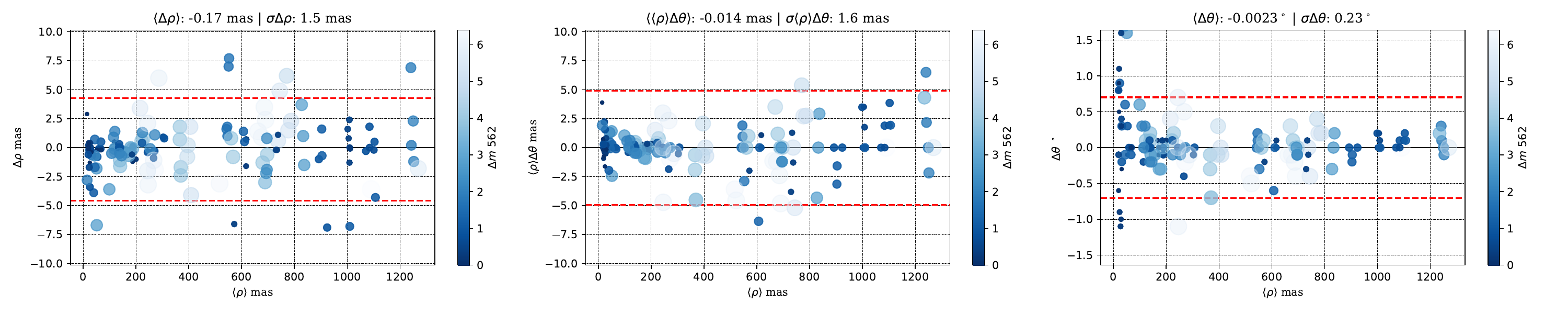}
\plotone{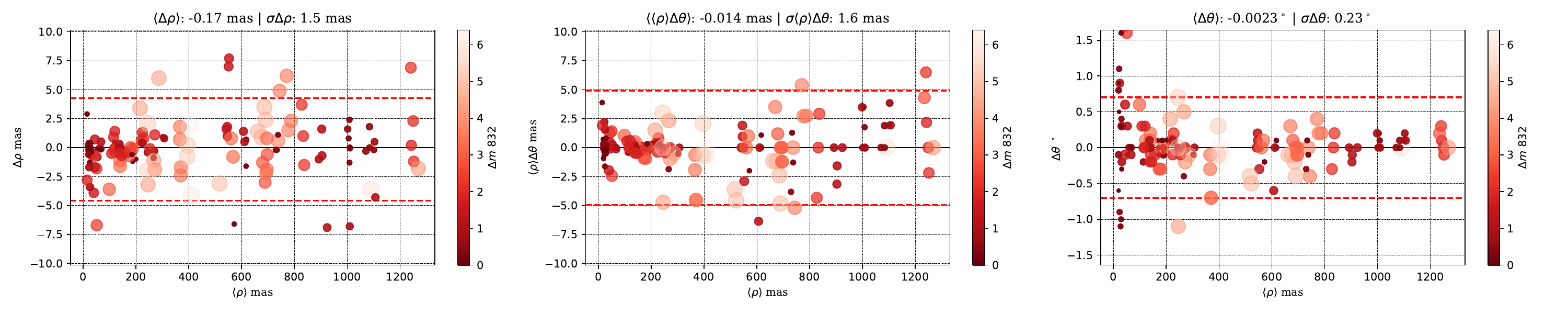}
\caption{Positional precision: residuals (ordinate) {\it vs.} $\rho$ (abscissa). Differences between the measured $\rho$ in the radial and tangent direction, $\Delta \rho$, $\overline{\rho} \cdot \Delta \theta$, both in mas, as well as in position angle $\Delta \theta$ (in \degr), as a function of the mean $\rho$ at a given epoch, for all the objects in our sample that have simultaneous observations in both filters. The first two rows of panels show the residuals considering the whole sample of binaries: first row for the blue filter and second row for the red filter. The third  and fourth rows show the effect of applying a 3$\sigma$-clipping to the data in order to compute the rms, for the blue and red filters, respectively. In these later plots, the dashed lines indicate the 3$\sigma$ rms. In these figures we have added a color coding to show the magnitude difference between the components, which adds a third dimension to the plots. Smaller, high-contrast, dots depict smaller $\Delta m$, while larger low-contrast dots the opposite. Blue dots were used for the blue filter plots, and red dots for the red filter plots. At the top of each figure, we include the mean residual value and its rms in the units of the corresponding ordinate.\label{fig:precision1}}
\end{figure}

In Figure~\ref{fig:precision1} we show, in three columns, the difference between the measured positions in the red and blue filters in the radial and tangent direction, $\Delta \rho$ (in the sense $\rho_{562}-\rho_{832}$), $\overline{\rho} \cdot \Delta \theta$, both in mas, as well as in position angle $\Delta \theta$ (in degrees, in the sense $\theta_{562}-\theta_{832}$), as a function of the mean $\rho$ at a given epoch, for all the objects in our sample that have simultaneous observations in both filters. The first two rows of panels show the residuals considering the whole sample of binaries: In the first (second) row the color-code (on the right to each figure) indicates the brightness difference in the blue (red) filter ($\Delta m$ hereafter).
The third and fourth rows show the effect of applying a 3$\sigma$-clipping to the data in order to compute the rms. In these later plots, the dashed lines indicate the 3$\sigma$~rms. The color coding in all plots shows the magnitude difference between the components, which adds a third dimension to the plots. Smaller, high-contrast, dots depict smaller $\Delta m$, while larger low-contrast dots the opposite. Blue dots were used for the blue contrast plots, and red dots for the red contrast plots.

For the whole sample shown, comprising 154 simultaneous observations in the blue and red channels for 50 unique binaries, the overall rms using an iterative $3 \sigma$ clipping estimate (lower two rows of Figure~\ref{fig:precision1}) indicate uncertainties of (1.5\,mas, 1.6\,mas, 0\fdg23 in ($\Delta \rho$, $\overline{\rho} \cdot \Delta \theta$, $\Delta \theta$) respectively, which can be considered as a representative uncertainty of the bulk of our observations. When computing the rms without a $\sigma$-clipping, the uncertainties are however substantially larger, indicating that there are some relevant outliers that should be explained.

Regarding the outliers seen in Figure~\ref{fig:precision1}, it seems clear that there are no systematic effects as a function of $\rho$, except that the largest residuals occur mostly on those objects that exhibit the largest $\Delta m$ between the components. On the other hand, there is an obviously larger scatter in $\theta$ at small $\rho$ (see also Figure~\ref{fig:precision2}).

The worst offender in $\rho$ is one observation of the large-separation binary ($\sim$0\farcs8) HIP~38635 (YSC~198~AaAb) at epoch 2023.01, with a huge blue-red difference of 21.3\,mas. This target was observed on other five epochs (see Table~\ref{tab:double}), with no difficulties. We have examined in detail the binary solution for this target on all epochs, and there are no problems with it, albeit the $\chi^2$ for its solution at the offending epoch is not particularly good: near 7 in the blue channel, and near 20 in the red channel; this having used a nearby PSF star. We note that the blue image is heavily distorted by AD, despite the fact that the zenith distance of this frame was only 30\degr. In fact, there is an observation at a larger zenith distance (47\degr) which has smaller residuals (4.9\,mas). The observation with the smallest residual (0.6\,mas) of this set has a zenith distance of 29\degr. The seeing on the night when this observation was acquired was good, with quartiles of 0\farcs38, 0\farcs42 and 0\farcs45. At the moment we do not have an explanation for the large difference between the blue and red measurements of this particular target. We note however the rather large magnitude difference of about 5.0\,mag in the blue, and 3.5\,mag in the red.

For the other targets that exhibit large residuals, we could find a variety of reasonable explanations: For example, for HIP~88937 (HDS~2560~AaAb, 14.6\,mas residual at epoch 2022.21), the companion is near the edge of the FOV ($\rho=$~1\farcs24); in one case even outside the FOV in the blue channel. Then, HIP~101966 (CVN~17AaAb, 14.3\,mas residual at epoch 2021.7141) exhibits a very noisy blue channel detection, with a large $\Delta m_{562} \sim 6.0$, in fact the fit to the binary was "guided" by the red channel ACF. Uncomfortably, the next two high-residual objects are both astrometric calibration binaries: the observation at epoch 2023.41 for HIP~67819 (HWE~28AB), with a $\rho$ of 1\farcs0 and small $\Delta m$, exhibits a residual of 10.9\,mas (another eight epochs exhibit a small residual), and HD~92015 (RST~3708) with a $\rho$ of 0\farcs55, a moderate $\Delta m$, with the next two high residuals at epoch 2023.17 with 7.7\,mas, and at epoch 2022.20 with a residual of 7.0\,mas (another three observations, at 2020.20, 2021.16, and 2023.17 do not show large residuals). 

Considering that HWE~28AB and RST~3708 are calibrators, and that both show large residuals in the 2023A run, one may be tempted to culprit the blue/red scale correction on that particular run. However, in this run two other astrometric standards were involved, OL~18 with $\Delta_\rho$ = 0.5\,mas (at $\rho=$ 1\farcs1), and RST~2368 with $\Delta_\rho$=1.5\,mas and $\rho$=0\farcs84. So, the scale correction cannot be the culprit, because it is applied equally to all images. Note that, as further discussed below, at $\rho$ larger than $\sim 0\farcs4$ the internal precision seems to be worse by a factor of two than at smaller $\rho$ (see Figure~\ref{fig:precision2}), and $3\sigma$ excursions (implying residuals of 6.0\,mas) are  not statistically impossible.

Regarding the outliers in $\theta$, it is obvious from the panels in the third column of Figures~\ref{fig:precision1} and~\ref{fig:precision2}, that these are systems with very small $\rho$, close to the diffraction limit, where the binary is only detected typically by one pair of widely separated fringes (see, e.g., Figure~\ref{fig:spectra}). The worst offender in this case is HIP~74165, resolved only on epoch 2021.16, with a $\rho$ of 15\,mas, and a \(\theta\) difference of +15\degr. Given its northern declination of +14\degr, it was acquired at a large zenith distance of 45\degr, so the blue image is very elongated which makes the binary fit uncertain. The next case is HIP~101472, with a similar $\rho$, and a \(\theta\) difference of +7\degr for the measurements on 2022.37. In this case, while the observations were all done at small zenith distance (5\degr) the larger magnitude difference (see Table~\ref{tab:double}) and small $\rho$ means that the fringe contrast is very weak, leading to a somewhat uncertain fit. Then follows HIP~59426, in the inner binary (S~634~Aa,Ab) of this triple system which at epoch 2020.04 and a $\rho$ of 23\,mas shows a \(\theta\) difference of +5\fdg5 (it has also another outlier at epoch 2023.50 and \(\theta\) difference of -3\fdg7, the second resolution of this system, see Table~\ref{tab:double})). The external Aa,Ac sub-system was detected only in the R channel. Accordingly, we performed a binary fit to the Aa,Ab sub-system in the blue frames, and a triple fit to the Aa,Ab plus Aa,Ac sub-systems in the red frames (see Table~\ref{tab:double}). Therefore, this large \(\theta\) difference could be caused by the different processing. To check on this suspicion, we have re-processed the system with the binary code alone for both filters. The mean difference in the \(\theta\) between the binary- and triple-fit for Aa,Ab is 0\fdg5 (average over five independent data cubes) so this can not be responsible for the large residual. The observations were done at relatively small zenith distance (16\degr) with rather moderate elongation of the ACF of the blue channel data, so the reason for the large residual is not obvious, apart from the fact that the fringes had a small contrast and the $\chi^2$ of the fits were not optimal (about 20). The last significant outlier in \(\theta\) is HIP~48333, at also small $\rho$ (25\,mas), with a \(\theta\) difference of -3\fdg7 at epoch 2020.02. There is a small but noticeable elongation on the blue channel images despite a small zenith distance (11\degr), but all fits have a reasonable $\chi^2 \sim 2.5$, albeit only one pair of fringes are available for fitting due to the small $\rho$.

One important outcome that can be appreciated from Figure~\ref{fig:precision1} is that the mean value of the differences in both $\rho$ and $\theta$ are well centered on zero (within the rms), which means that our Red-to-Blue scale correction, discussed in Section~\ref{sec:matchrb}, has been properly applied in all epochs and that there are no evident systematic offsets neither in $\rho$ nor in $\theta$. Overall, this means that the red and blue measurements can be safely considered as independent measurements of the same quantity and used as such in our orbit fitting later on. Furthermore, by comparing the panels in the first and second columns, we see that the rms of the radial and tangential directions are equivalent.

\begin{figure}[ht!]
\epsscale{1.1}
\plotone{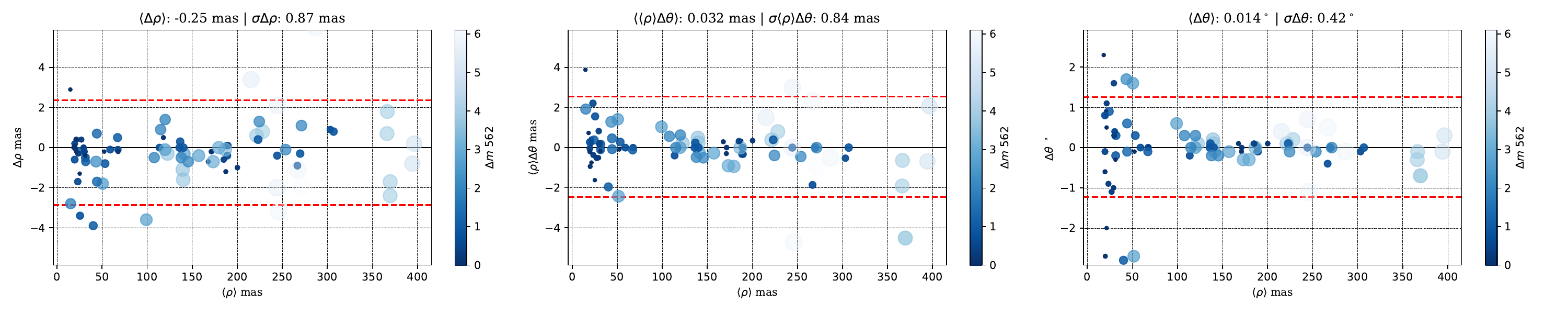}
\plotone{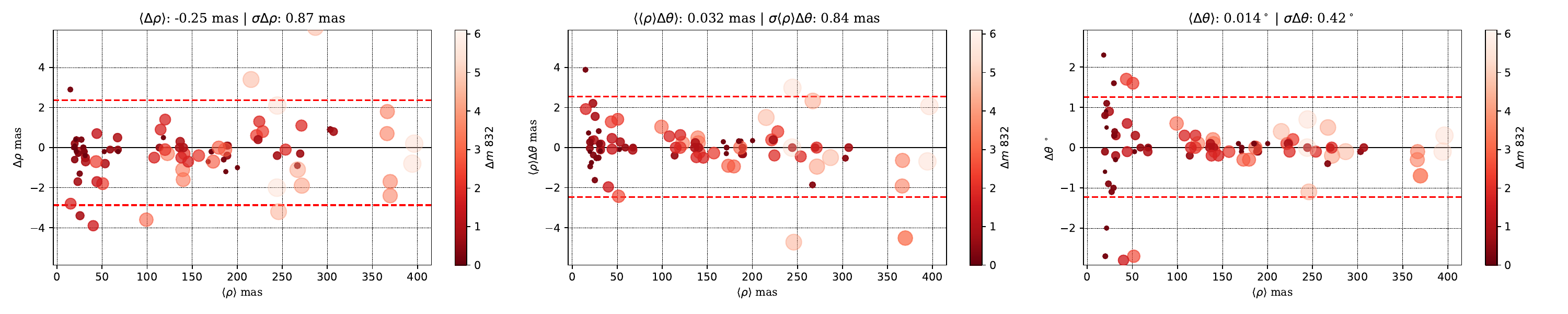}
\plotone{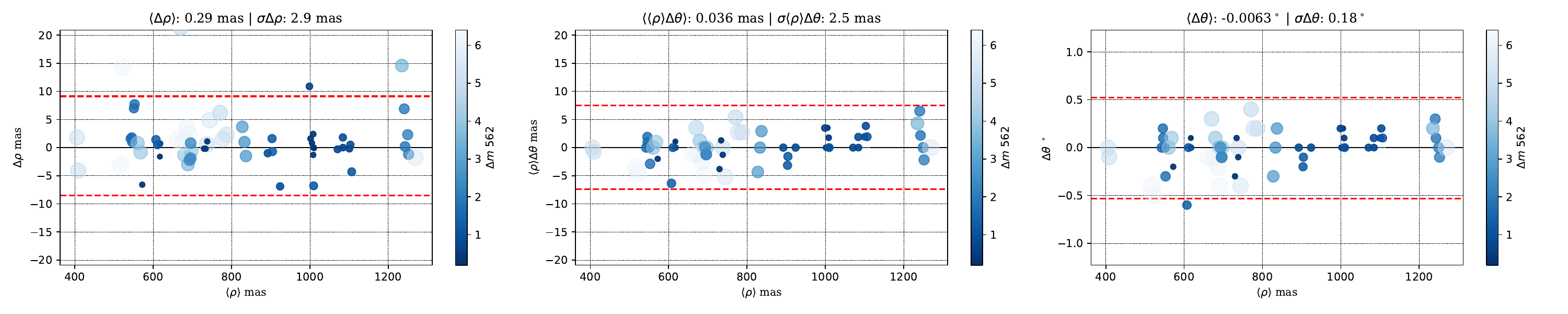}
\plotone{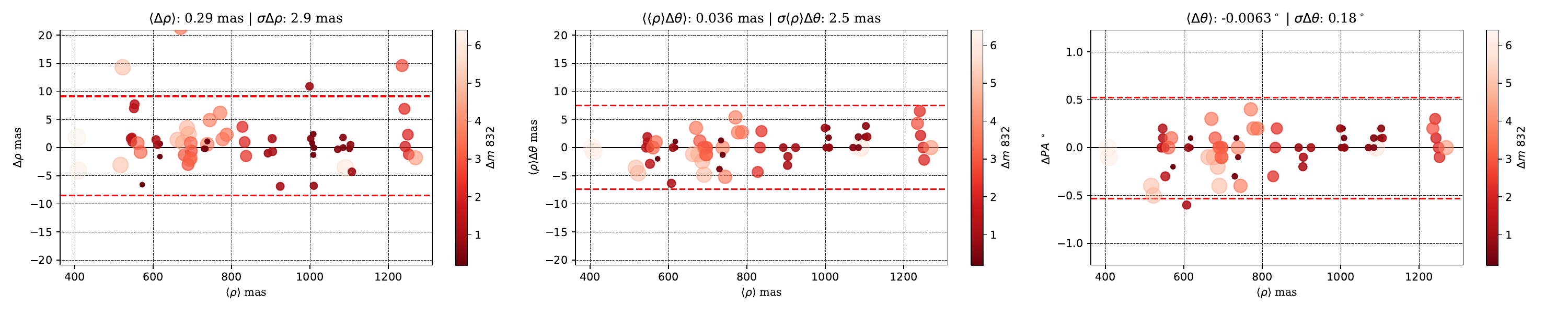}
\caption{The same general description for rows three and four of Figure~\ref{fig:precision1} are valid  for this figure. Here, however, rows one and two are for the range with $\rho<$0\farcs4 (program stars), while rows three and four are for the larger $\rho$ binaries (mostly astrometric standards).}\label{fig:precision2}
\end{figure}

It is also apparent from Figure~\ref{fig:precision1}  that the overall scatter is somewhat dependent on $\rho$. For example, it appears that the scatter in the radial and tangent directions suffers a change near a $\rho$ of 0\farcs4. Several of the large-separation binaries are actually astrometric calibration binaries (see Table~\ref{tab:double} and Section~\ref{sec:abscal}), while objects at closer $\rho$ are program stars. This is indeed confirmed by examining Figure~\ref{fig:precision2}. At small $\rho$, we find that the $\rho$ measurement errors would usually surpass 1\,mas (e.g., in $\rho$ a 0.87\,mas rms of the $\rho$ difference should imply an uncertainty of $0.87/\sqrt{2}$ per measurement), while at large $\rho$ the per-measurement uncertainty would be more like 2\,mas (see also Table~\ref{tab:astromcomp}). In terms of PA, we see the opposite trend; at small $\rho$, the scatter is larger than at large $\rho$. This is evident in Figure~\ref{fig:precision2}, which is due to the much larger scatter in \(\theta\) at $\rho$ close to the diffraction limit (already mentioned when commenting on in Figure~\ref{fig:precision1}). Interestingly, this large rms in \(\theta\) does not translate into a larger rms in the tangential direction, which exhibits an rms of 0.84\,mas similar to that in the radial direction, due to the very small $\rho$ of these binaries (compare the middle and right plots on rows one and two of Figure~\ref{fig:precision2}). We note that our pipeline does not correct for differential distortion between channels, nor for possible scale differences in the X and Y axes, as was done by \citet{2009AJ....137.5057H} for the Differential Speckle Survey Instrument (see their Equation~(3) on Section~3). It is likely that the intrinsic Zorro optical design could be the main reason for the apparent increase in uncertainty at larger $\rho$. Another factor that increases the error for larger separation is the atmospheric differential tilt discussed in \citet{2022AJ....164...58T} (see Equation~(19) on Section~2.4).


In terms of the repeatability of the relative photometry, Figure~\ref{fig:photom1} shows the scatter of the repeated measurements over all available epochs, as a function of mean magnitude difference and mean $\rho$, for both filters (rows one and three). A straight $3 \sigma$ clipping estimate for the whole sample shown indicates uncertainties of 0.091\,mag for both filters (rows two and four). This figure also includes a third dimension:  mean $\rho$ in rows one and two, and mean magnitude difference in rows three and four. The color coding in this figure is analogous to that of Figure~\ref{fig:precision1} and Figure~\ref{fig:precision2}. In Figure~\ref{fig:photom1} the ordinate $\Delta m - <\Delta m>$ is the residual of a particular $\Delta m$ measurement with respect to the mean of all measurements (over all epochs) in a given filter for that target, this is termed $\Delta \Delta m$ below.

As was the case for the positional uncertainties, some outliers are worth noticing. The  tighter component (Tok~722AaAb, $\rho$ of 20\,mas) of the triple system HIP~63377 shows in one of its measurements (epoch 2022.37) a very large $\Delta m \sim 1.6$ which is not compatible with previous or posterior observations, which indicate a very small $\Delta  m$. This sub-system could only be detected in the red channel. We re-processed this sub-system with the binary code instead of the triple-fitting code, and we obtained a $\Delta m = 0.0$, compatible with the other observations. We also note (see Table~\ref{tab:double}) that its $\rho$ at this epoch is rather discrepant with the values before and after. All the epochs were acquired with a similar zenith distance of $\sim 34$\degr, and the seeing was commensurable, so there is no obvious reason for the discrepancy. Variability of one of the components could be perhaps advocated, but this does not explain the difference in $\rho$, and, besides, it is rather large (1.6\,mag). Another outlier is HIP~88937; an astrometric binary used earlier on in the program, and later deprecated. It has a large $\rho$ of nearly 1\farcs3, which puts the companion very near the edge of the FOV, hence some of its flux is lost making the photometry unreliable. The last outlier worth noting is HIP~101472 at epoch 2022.37, with a difference of $\Delta \Delta m = 0.73)$. This is the same outlier mentioned earlier, with a large difference in $\theta$, and likely for the same reasons.



Apart from the outliers mentioned above, there are no obvious trends as a function of either magnitude contrast or $\rho$. Given the difference in positional precision for small and large $\rho$ noted before, we examined the photometric precision in the same $\rho$ ranges, namely $[0.0,0.4)$\arcsec and above 0\farcs4, which is shown in Figure~\ref{fig:photom2}. At large $\rho$ (bottom row) the photometric repeatability seems a bit worse, as already noticed in Section~\ref{sec:binproc}. At small $\rho$ (top row), the red filter seems to be a bit better than the blue one, even considering that the more difficult-to-measure tighter systems are only resolved in that filter (hence the larger number of points at very small $\rho$).

\begin{figure}[ht!]
\epsscale{1.1}
\plotone{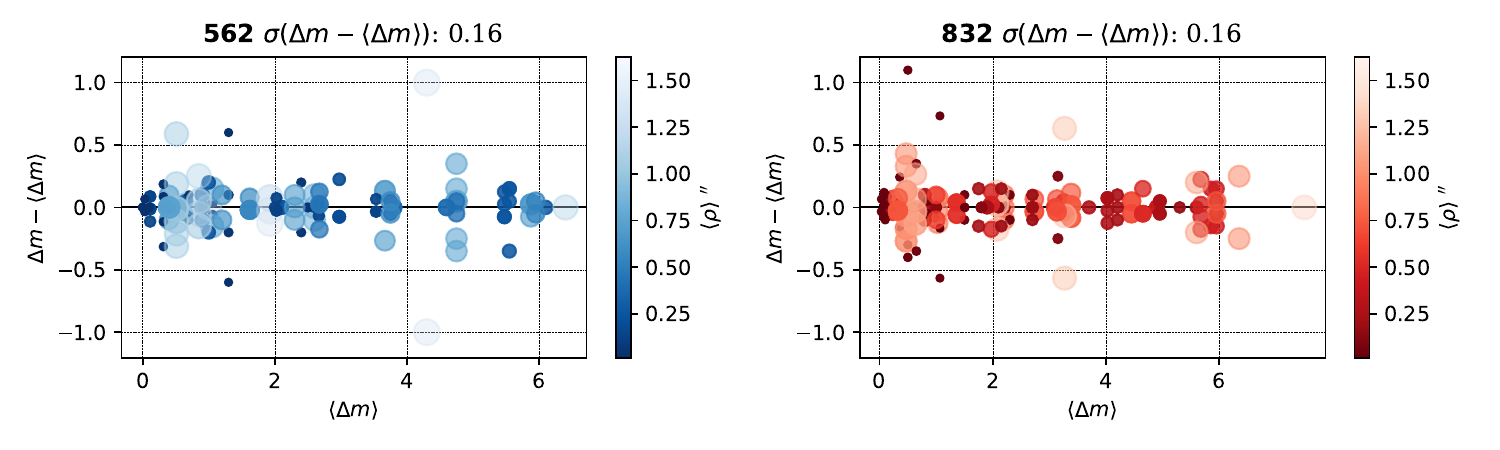}
\plotone{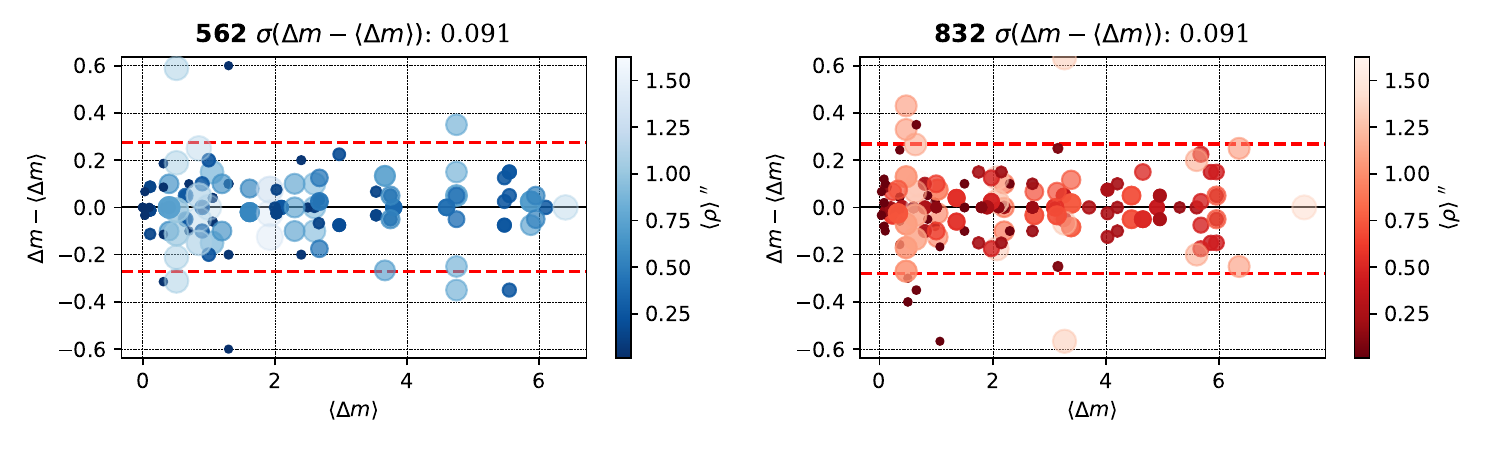}
\plotone{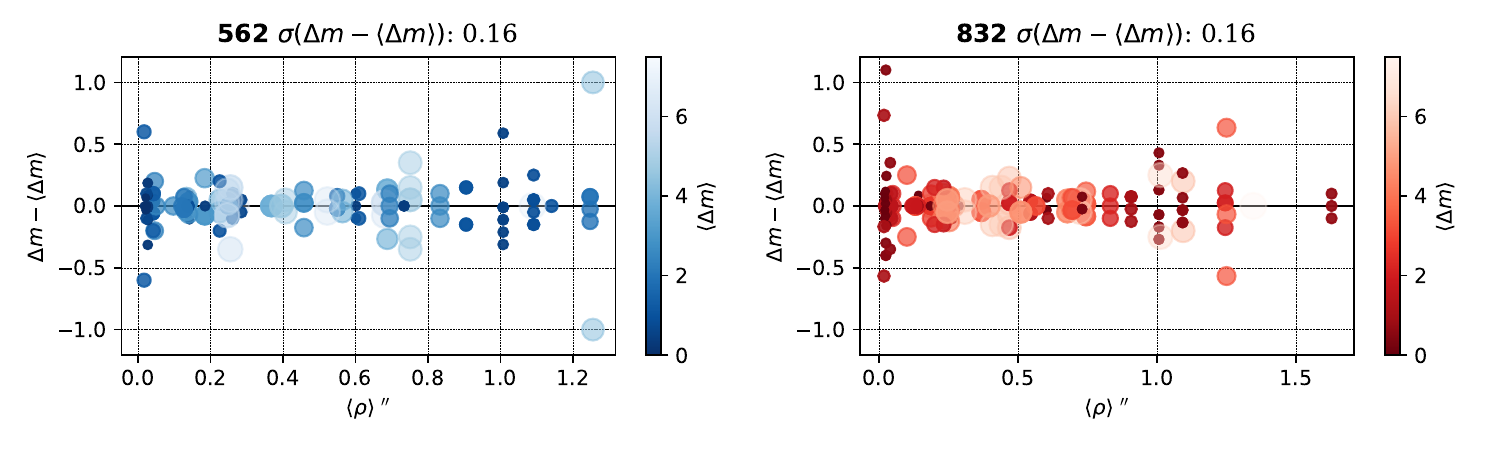}
\plotone{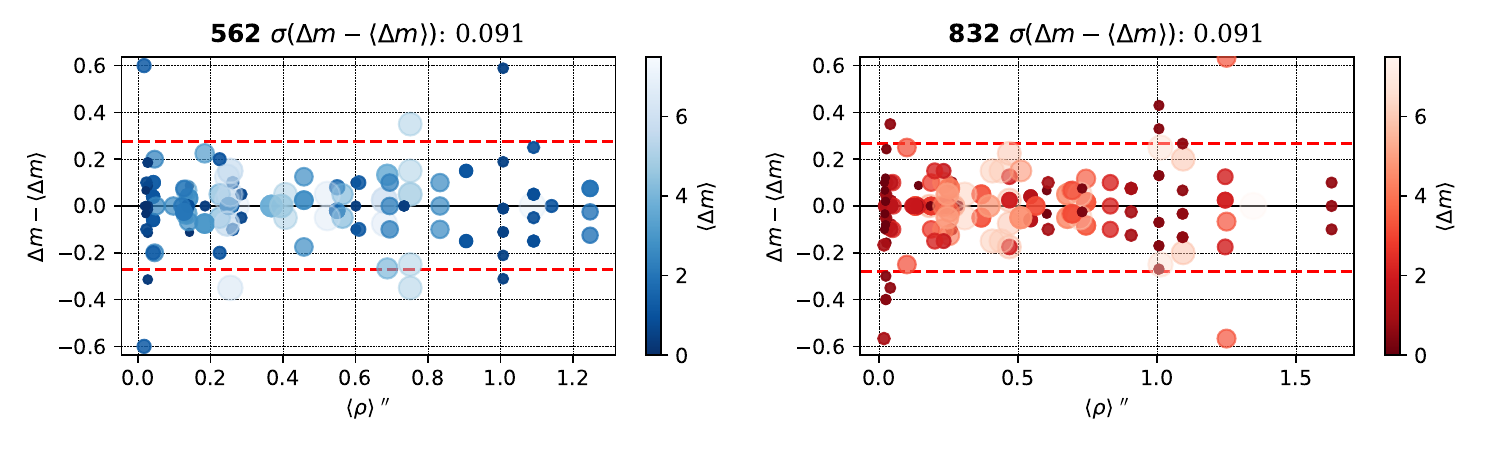}
\caption{Photometric precision: residuals (ordinate) {\it vs.} contrast (abscissa) and $\rho$ for the 562\,nm (left column) and 832\,nm (right column) filters for all our binaries with observations in the two filters. The first row is for the whole sample, while in the second row  applies a 3$\sigma$-clipping to compute the rms; in this case, the dotted lines indicate the 3$\sigma$ rms. The third and fourth rows are similar, but as a function of mean $\rho$ and magnitude difference. The color coding in this figure is analogous to that of Figure~\ref{fig:precision1} and Figure~\ref{fig:precision2}. At the top of each figure, we include the mean residual value and its rms in the units of the corresponding ordinate.}\label{fig:photom1}
\end{figure}

\begin{figure}[ht!]
\epsscale{1.1}
\plotone{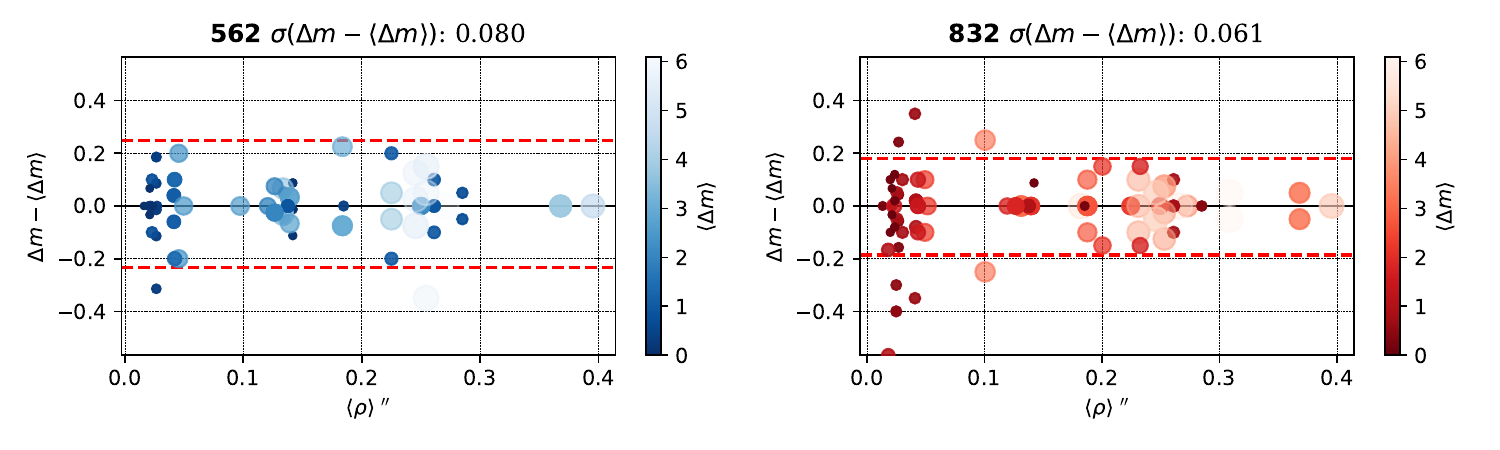}
\plotone{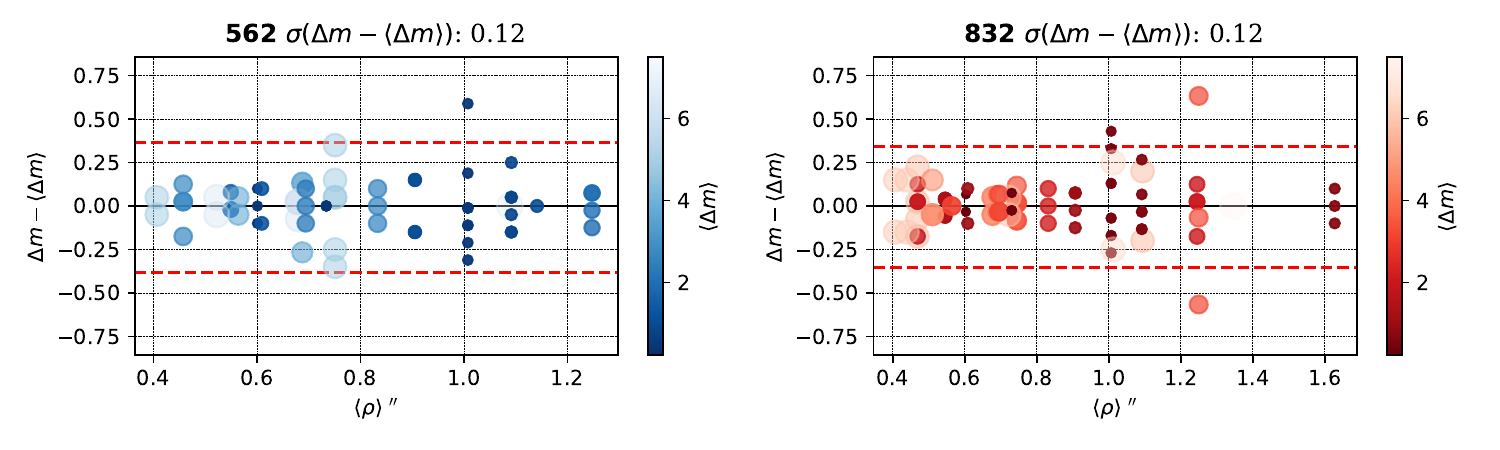}
\caption{Same as row four on Figure~\ref{fig:photom1}, except that the top row is for the range with $\rho< 0\farcs4$ (program stars), while the bottom row is for the larger $\rho$ binaries (mostly astrometric standards).}\label{fig:photom2}
\end{figure}


\section{Results and some preliminary orbits} \label{sec:orbits}

In Table~\ref{tab:double} we present our measurements for the systems that could be resolved. In the first column we list the WDS Identification, in the
second column the Discoverer Designation (DD) from WDS\footnote{Objects with HD or HIP number in the DD means that they are new binaries, and a DD name is pending from the WDS curators} and in the third column is the HIPPARCOS number (when available). The fourth column gives the epoch in fraction of Julian yr minus 2000.0, the fifth column indicates the filter name and the sixth column gives the number of cubes (of 1000 images each) averaged to produce our measurement. We then give (columns 8, 9 and 10): $\theta$ in degrees, the tangential formal error in mas (from the binary or trinary fit code), the $\rho$ in arcsec, and the formal uncertainty in mas (from the binary or trinary fit code). The eleventh column gives the magnitude contrast in the respective filter and in column twelve a code (q means firm quadrant detection on the SAA images, : means uncertain measurement, data are noisy and $\Delta$m is likely over-estimated, or tentative resolution). If there is an orbit published for the target, $\Delta\theta$ (in degrees) and $\Delta\rho$ (in mas) indicate the residual of our measurement with respect to the predicted \(\theta\) and \(\rho\) computed for the epoch of observation, in this case we give the orbit reference from Orb6\footnote{See \url{http://www.astro.gsu.edu/wds/orb6.html}.}. The last column gives some relevant comments, when applicable. An overview of the data contents of this table is presented in Figures~\ref{fig:histsep}, ~\ref{fig:histmag} and~~\ref{fig:sepmag}. We note that there are six (nine) objects with $\Delta m \ge 5.5$ in the blue (red) filter. Of these, only one has $\Delta m \ge 5.5$ in both filters (Tok~791AB), and of the nine objects with large $\Delta m$ in the red filter, only two of them have measured $\Delta m$ in both filters (Tok~791AB and HD~100378). In terms of $\rho$, eleven objects have $\rho$ smaller than 25\,mas.
%
%
%
%
\FloatBarrier
\startlongtable
\begin{deluxetable}{cccccccccc}
\tablecaption{Unresolved targets.\label{tab:single}}
\tablecolumns{10}
\tablewidth{0pt}
\tabletypesize{\tiny}
\tablehead{\colhead{WDS} &
\colhead{DD} &
\colhead{HIP} &
\colhead{Epoch} &
\colhead{Filt} &
\colhead{Ncub} &
\colhead{MinSep} &
\colhead{Det@0.15} &
\colhead{Det@1.00} &
\colhead{Comments} \\
\colhead{} &
\colhead{} &
\colhead{} &
\colhead{+2000.0~yr} &
\colhead{} &
\colhead{} &
\colhead{\arcsec} &
\colhead{mag} &
\colhead{mag} &
\colhead{}
}
\startdata
00466$-$5406 & HD 4520       &       3640   &  19.5438 & 562 &  3 &  0.014  & 7.86  &  7.01 & \\  
           &               &              &  19.5438 & 832 &  3 &  0.021  & 8.43  &  7.66 & \\
\hline
00507$-$5059 & HD 4919       &       3949   &  19.5438 & 562 &  3 &  0.014  & 7.54  &  7.09 & \\  
           &               &              &  19.5438 & 832 &  3 &  0.021  & 8.07  &  7.46 & \\  
\hline
00527$-$2400 & HD 5445       &       4104   &  19.5438 & 562 &  3 &  0.014  & 8.47  &  6.64 & \\  
           &               &              &  19.5438 & 832 &  3 &  0.021  & 6.05  &  6.83 & \\  
\hline
01004+1714 & HD 5873       &       4696   &  20.8246 & 562 &  3 &  0.014  & 0.00  &  4.65 & \\  
           &               &              &  20.8246 & 832 &  3 &  0.021  & 6.10  &  5.53 & \\
\enddata
\end{deluxetable}
\FloatBarrier

\begin{figure}[ht]
\centerline{
\includegraphics[width=13cm]{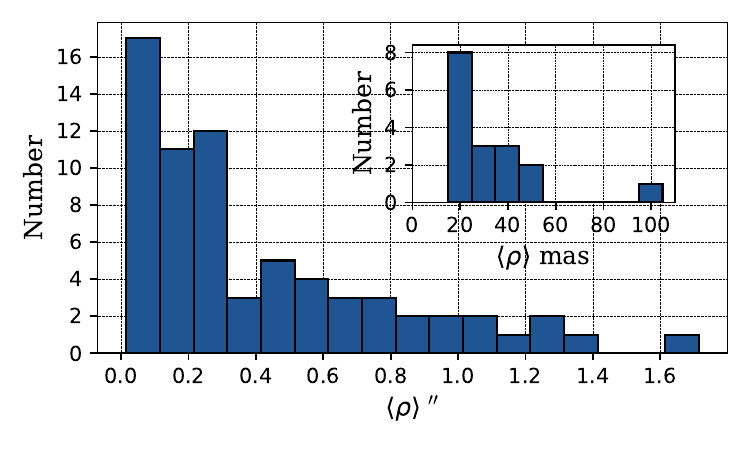}
}
\caption{Histogram of mean $\rho$ in bins of 0\farcs1, computed over all epochs and all filters, for all the resolutions presented in Table~\ref{tab:double}. The inlet shows the histogram for objects with $\rho$ smaller than 100\,mas in bins of 10\,mas.}
\label{fig:histsep}
\end{figure}

\begin{figure}[ht]
\centerline{
\includegraphics[width=17cm]{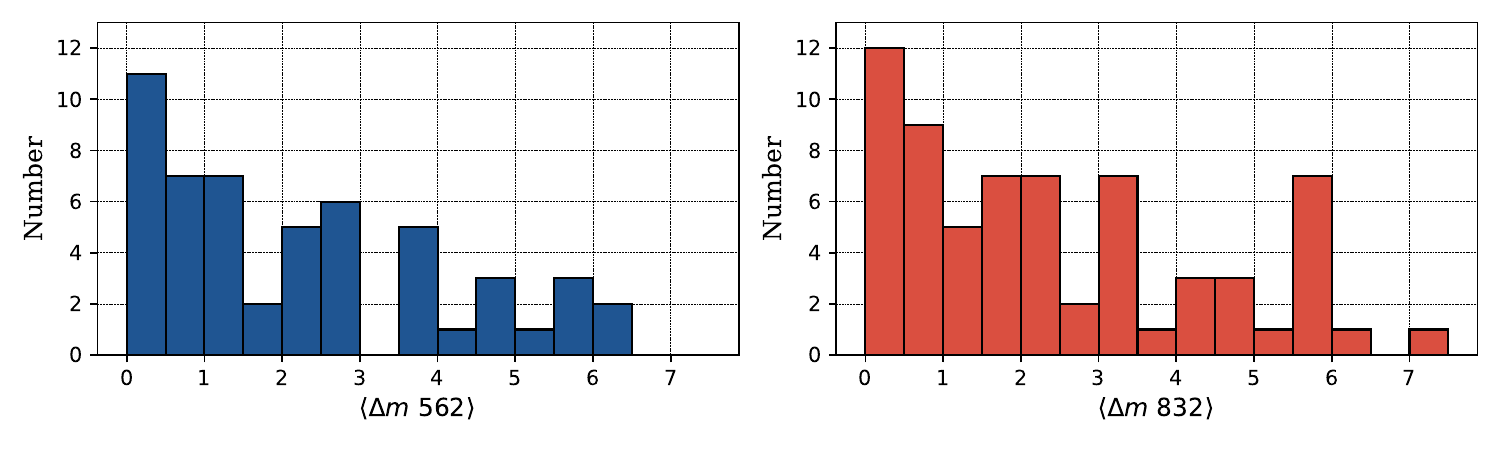} 
}
\caption{Histogram of mean $\Delta m$ in bins of 0.5\,mag, computed over all epochs for the blue filter (left) and the red filter (right), for all the resolutions presented in Table~\ref{tab:double}.}
\label{fig:histmag}
\end{figure}

\begin{figure}[ht]
\centerline{
\includegraphics[width=17cm]{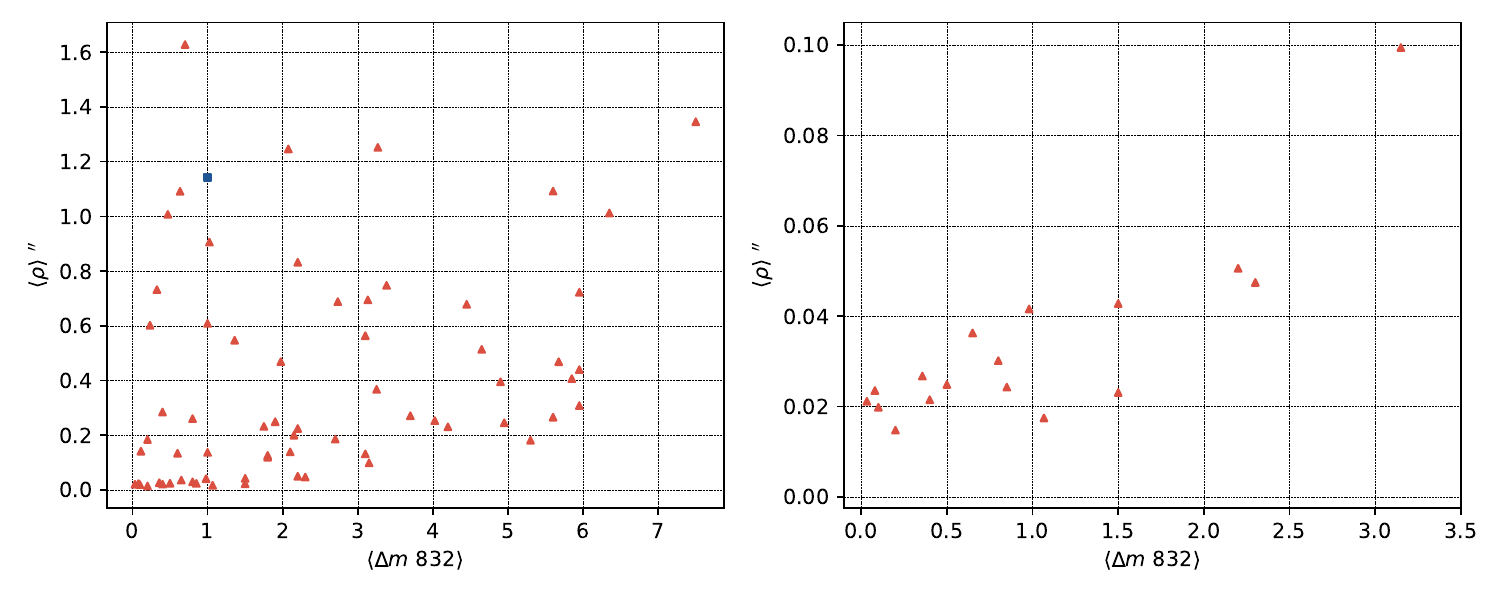} 
}
\caption{Average $\rho$ of the components (over all epochs) as a function of the mean $\Delta m$ in the red filter for all the resolutions presented in Table~\ref{tab:double}. The left panel is for the whole sample, while the right panel is for objects with $\rho$ less than 100\,mas. In the left panel, the blue square indicates the only object with no detection in the red filter (STF~1998AB), for this we plot the blue filter contrast instead.}
\label{fig:sepmag}
\end{figure}

In Table~\ref{tab:single} we present the list of unresolved targets. Here we also include PSF objects that were assumed to be single-stars, so it is no surprise that most of them are included in this table\footnote{because they are not listed in the WDS, they do not have a DD name, instead we adopt the HD or HIP number for them.}. The meaning of the first six columns in this table is the same as that of Table~\ref{tab:double}. The seventh column gives the angular resolution in arcsec, while the eighth and ninth columns give the 5$\sigma$ magnitude difference -the detection limits of possible companions- at 0.15 and 1.00\arcsec respectively. Four PSF stars were serendipitously found to be new binaries, so they were moved to Table~\ref{tab:double}.

\subsection{Preliminary orbits}
\label{sec:orb}

With the available Zorro measurements, it was possible to calculate tentative orbits for some targets. However, we must emphasize that in most cases a follow-up will be required to complete the orbital coverage. Orbits derived from the present data should be considered preliminary; they are presented here only with the purpose of evaluating Zorro's capabilities.  As a strategy, all pairs that could possibly be observed and followed up with smaller facilities have been moved to SOAR, while all very tight and/or faint companions will be continued to observe with Zorro@GS.

We note that, in Table~\ref{tab:double}, most objects with large $\rho$ (larger than about 0\farcs4) are calibration binaries, while tighter systems are our program targets. All objects in this table with the comment "Was PSF star" means they are newly confirmed binaries (four objects in total) with red companions of various $\rho$ and a relatively large $\Delta m$. The most extreme case is HIP 82216, with a companion having $\Delta m \sim 6.0$ in the 832 filter and undetected in the 562 filter. There are also seven other new binaries which are either very tight (e.g., HIP~74165 and HIP~101472) or have large $\Delta m$ (e.g. HIP~5146 and HIP~19120); mostly detected in the red channel only.\\


\begin{figure}[ht]
\centerline{
\raisebox{11mm}{\includegraphics[width=8.5cm]{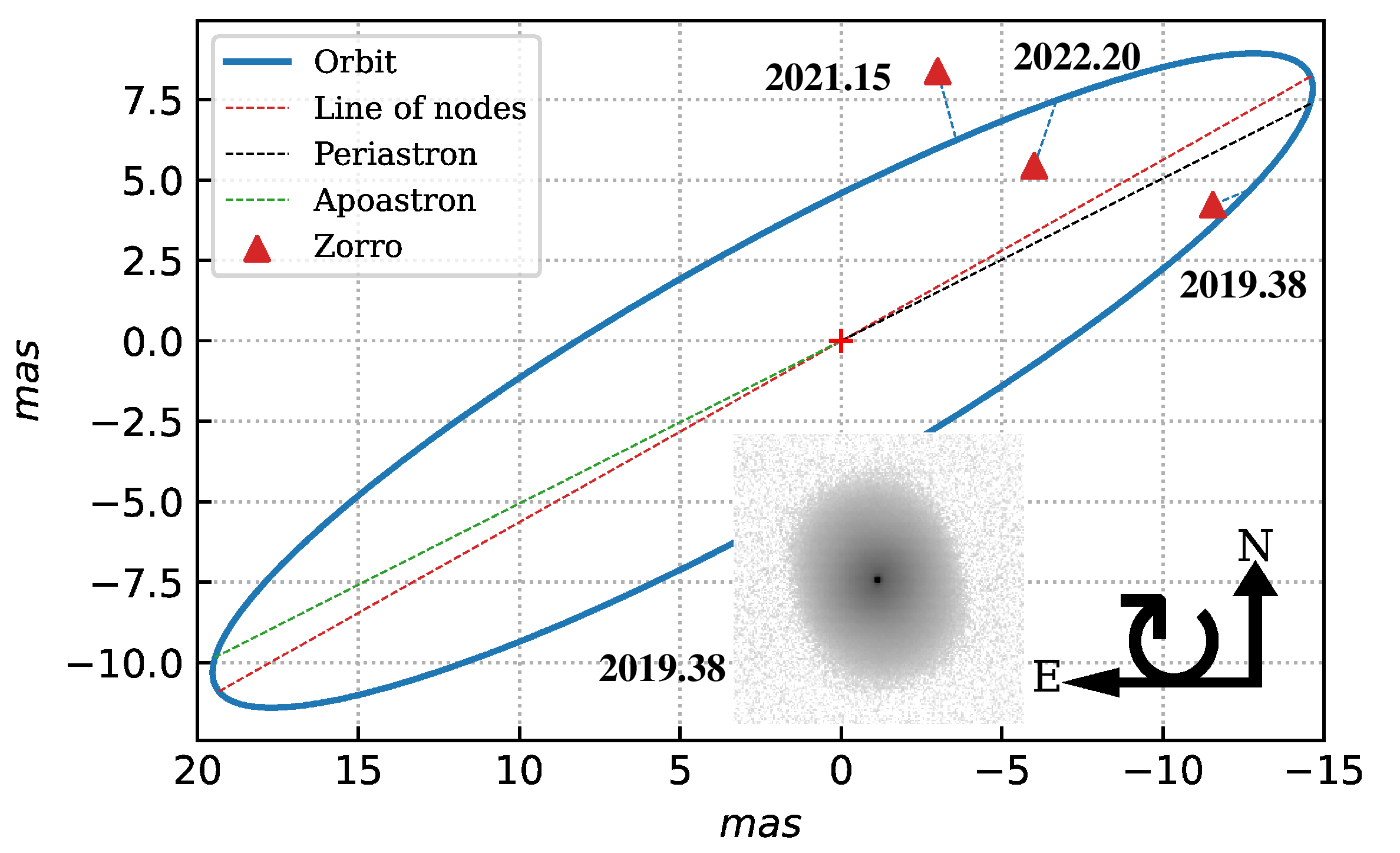}} 
\includegraphics[width=8.5cm]{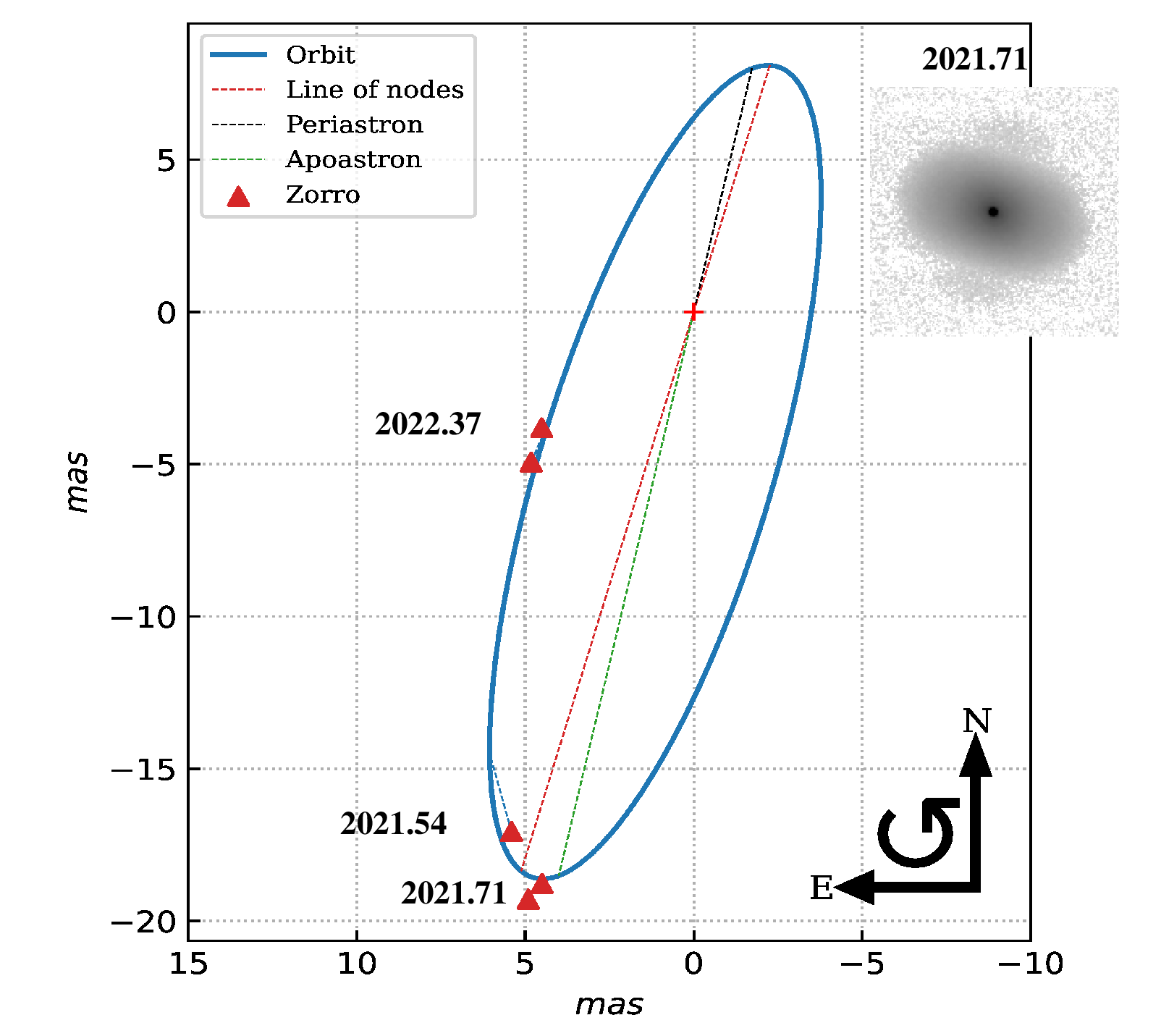}
}
\caption{Preliminary visual orbits for HIP~76400 Aa1,Aa2 (left, P=368.6\,days, a=19.5\,mas) and HIP~101472 Aa,Ab (right, P=354.9\,days, a=13.9\,mas) derived by joint fitting of Zorro positions and available RVs. The inserts show the Speckle power spectra recorded near maximum $\rho$ at 832\,nm, with the same orientation.
\label{fig:orb}
}
\end{figure}

Of particular interest are, of course, objects resolved for the first time. Figure~\ref{fig:orb} illustrates tentative combined spectro-interferometric orbits based on our Zorro data. Their visual orbital elements are listed in Table~\ref{tab:orbel}.

HIP~76400 (HD~139059, WDS 15362$-$0623, primary G6V) is a quadruple system composed of an internal double-line spectroscopic binary Aa1,Aa2 unresolved until now, with a companion Ab at 0.3\arcsec (Tok~301Aa,Ab), and another distant companion, B, at 80\arcsec (TOK 301AB). In our solution for Aa1,Aa2, the outer 70-yr orbit of Aa,Ab was fitted as well (it causes a substantial trend in the systemic velocities). For comparison, we also give the orbital parameters derived from the SB9 spectroscopic orbit of HIP~76400Aa1,Aa2.

HIP~101472 (HD~195719, WDS 20339$-$2710, primary G8V) is a triple system (possibly belonging to the Thick-Disk, \citet{2019AJ....158..222T}) composed of an internal double-line spectroscopic binary Aa,Ab not resolved until now, with a faint companion B at 53\arcsec (CBL~178) that exhibits very small motion, and has no published orbit (period estimated at 200~kyr). In Table~\ref{tab:orbel} we also show the (purely) astrometric orbit Aa,Ab from the non-single catalog produced by Gaia DR3 \citep{2022yCat.1357....0G}\footnote{The values for $\Omega$ and $\omega$ have been corrected for their 180~deg ambiguity in Gaia DR3, to match our joint RV/
astrometric solution}.



\begin{table}[ht]
\caption{Tentative orbital elements for HIP~76400 Aa1,Aa2 and HIP~101472 from a joint astrometric/spectroscopic solution.}
\begin{center}
\begin{tabular}{cccccccc}
\hline\hline
Pair & P & T & e & a & $\Omega$ & $\omega$ & i \\
& days & yr &  & mas & \degr & \degr & \degr \\
\hline
Hip 76400 Aa1,Aa2 & 368.6 & 2018.3 & 0.143 & 19.5 & 119.4 & 192.3 & 102.0 \\
SB9 & 368.51 & 2018.32 & 0.1370 & ---& --- & 192.0 & --- \\
\hline
Hip 101472 & 354.9 & 2016.4 & 0.397 & 13.9 & 344.4 & 12.2 & 74.1 \\
Gaia DR3 & 353.98 & --- & 0.410 & --- & 356.9 & 33.3 & 91.2 \\
\hline\hline
\end{tabular}
\end{center}
\label{tab:orbel}
\end{table}

Both systems have periods close to one yr, which makes it difficult to adequately sample the orbits, due to the yearly visibility cycles, fixed telescope schedules, and the need to observe as close to the meridian as possible.

These two cases are a good example of the limitations of the Zorro@GS setup. Their semi-major axes ($\sim$19\,mas and $\sim$15\,mas) are below the diffraction limit of Gemini at 832\,nm, so our tentative resolutions do not yield accurate measurements of the positions. Even if more data are accumulated, the prospect of measuring masses with sufficient (a few percent) accuracy for targets like these is unlikely. To constrain the inclination of these edge-on orbits, the pairs must be resolved at $\rho$ substantially less than the maximum, below the diffraction limit. A long-baseline interferometer like VLTI is needed in these cases.

In a forthcoming paper, we will publish orbits for other targets in our sample, combining them with new RV measurements acquired by our team with the FEROS Echelle spectrograph at the MPG 2.2\,m telescope at ESO/La Silla.\\

\subsection{Comments on individual objects} \label{sec:comments}

In this section, comments regarding multiplicity are mostly based on Tokovinin´s Multiple Star Catalogue (MSC\footnote{Updated version available at \url{http://www.ctio.noirlab.edu/~atokovin/stars/}}, \citet{1997A&AS..124...75T, 2018ApJS..235....6T}).\\

{\bf HIP~7869 = 01412$-$6741 = HD~10607} is a nearby quadruple system of 2+2 architecture. The outer 33\farcs5 common proper-motion pair (CPM hereafter) A,B is known as LDS~56; its estimated period is 76\,kyr. The 12.4\,mag secondary star B was resolved at SOAR as a 0\farcs49 pair; it also has double transits in Gaia, indicating a double source. The 8.32\,mag G0V primary star A has double lines \citep{2004A&A...418..989N}. It has been resolved by Zorro twice, but only in the red channel; in 2019.54 and 2020.90, at $\rho$ of 25.6 and 34.7~mas respectively, with a moderate $\Delta m \approx 0.8$ mag (Figure~\ref{fig:7869}); the pair was unresolved on two subsequent visits in 2021.72 and 2022.76 (see Table~\ref{tab:double}). The estimated period of Aa,Ab is 3\,yr. Both A and B are independently detected as astrometric binaries by Gaia DR3 (RUWE 8.5 and 5.2), so their parallaxes 
(12.67 and 13.04 mas) are unreliable. The masses range from 0.4 to 1.1 M$_{\odot}$. As we collect more epochs, a good prospect exists for a combined inner orbit.\\

\begin{figure}[ht]
\centerline{ \includegraphics[clip, trim=0 0 200 600, width=0.8\textwidth]{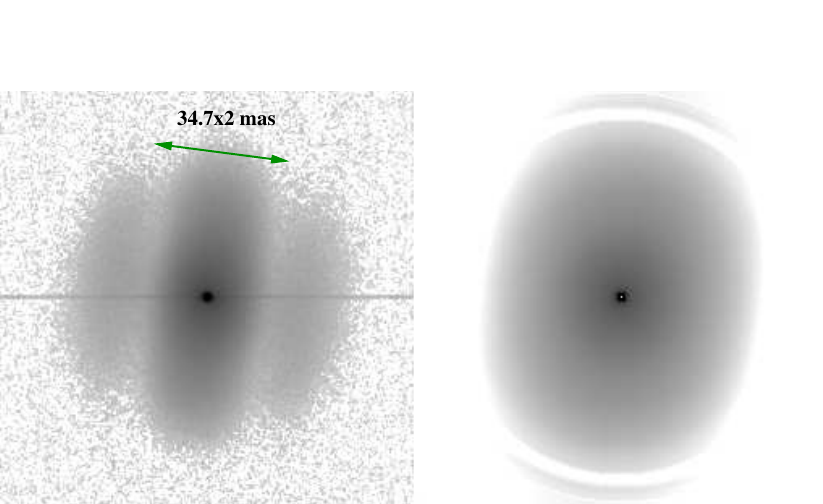} }
\caption{The power spectrum of HIP~7869 (component A) recorded on 2020.90 at 832\,nm (left), and its model accounting for the AD (right).}
\label{fig:7869}
\end{figure}

{\bf HIP 30953 = 06298$-$5014 = HD~46273.} This is a 2+2 quadruple system with an outer $\rho$ of 12\farcs2 (DUN~30). Both pairs A,B, and C,D have known visual orbits with periods of 53 and 99\,yr, respectively. The pair A,B (R~65) has been observed with Zorro eight times, covering the periastron of its eccentric ($e=0.97$) orbit in 2021.39. Double lines were expected near the periastron, but unfortunately no RVs were secured at this epoch. The residuals to the latest orbit\footnote{In Circular 207, available at \url{https://www.usc.gal/astro/circularing.html}.} (see also Table~\ref{tab:double}) indicates that it should be updated. Interestingly, pairs A,B and C,D move in opposite directions (inclinations 33 and 157 \degr), excluding their coplanarity. The mass ratios in both pairs are close to one, but the component's masses are different; 1.5 and 0.8~$M_\odot$. The dynamical parallaxes deduced from the estimated masses and orbits are 20.1 and 21.5~mas and agree with the Hipparcos parallax of 19.4\,mas. Gaia gives no parallaxes for these pairs. Interestingly, \citet{2007ApJ...658.1289T} found a 16\,AU debris disk around A.

{\bf HIP~38625 = 07546$-$0125 = HD 64606, K0V} is a triple  system at 20\,pc from the Sun (GJ~292.2). WDS lists two pairs, HDS1125AB at 4\farcs9 and YSC~198Aa,Ab at 0\farcs5 with similar magnitude differences of 4\,mag. However, the 4\farcs9 pair is spurious; it is not confirmed by Gaia and most likely results from the wrong interpretation of the double-star signal in the Hipparcos data reduction (a similar case was recently documented at SOAR). Star A is also a spectroscopic binary with $P=450$~days \citep{2002AJ....124.1144L}.  Its estimated semimajor axis is 66\,mas, but the RV amplitude implies a minimum mass ratio of 0.25, making it a poor candidate for speckle resolution. Nevertheless, the star was visited by Zorro five times, with $\rho$ decreasing from 0\farcs79 in 2020 to 0\farcs74 in 2022 and little change in $\theta$ (direct motion). So, this pair has opened up from 0\farcs52 in 2010, when it was resolved by \citet{2017AJ....153..212H}, and now is closing again. Its estimated period is 30\,yr, and a preliminary orbit can be fitted to the data (albeit on a very short arc). The estimated amplitude of the wobble caused by the subsystem is 13\,mas, so the speckle data should be fitted by two Keplerian orbits to determine the inclination of the subsystem Aa,Ab. The object has fast PM, large RV (102.2\,km\,s$^{-1}$), and is located slightly below the standard main sequence (metal-poor).

\bf HIP~55505 = 11221$-$2447 = HD 98800, K4V} according to SIMBAD, the primary is a T Tauri star TWA~4 in a quadruple system consisting of close spectroscopic pairs Aa,Ab and Ba,Bb on a 205\,yr orbit around each other. We have not resolved the inner pairs and measured the outer pair I~507AB which is closing down, approaching periastron.

\begin{figure}[ht]
\centerline{
\includegraphics[width=18cm]{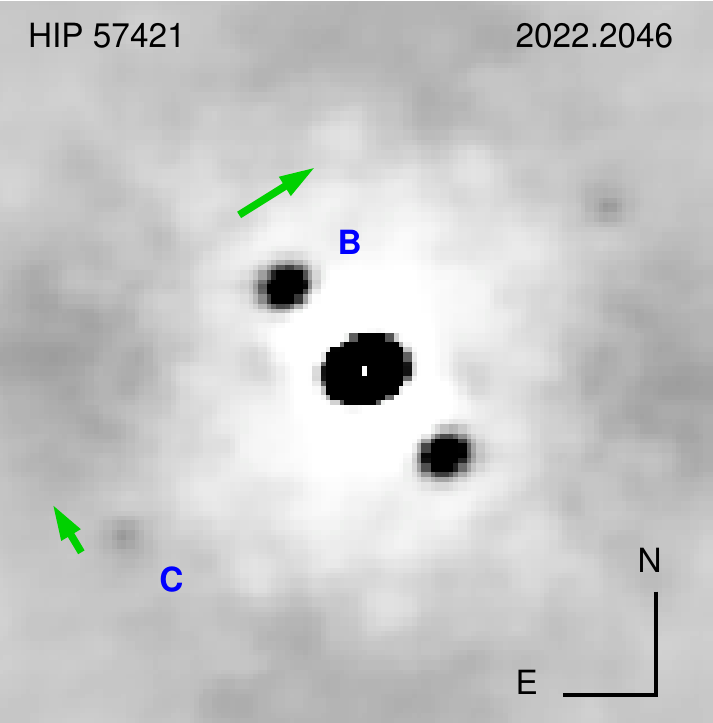} }
\caption{Speckle ACF of the triple system HIP~57421 recorded on 2022.20 at 823 nm  (in negative rendering). The inner 0.18\arcsec pair is A,B, and the faint tertiary C is seen at nearly orthogonal angle at 0\farcs46. Green arrows show the direction of motion. \label{fig:57241} 
}
\end{figure}

{\bf HIP 57421 = 11464$-$2758 = HD 102301, G0V} is a strange triple system with comparable $\rho$ between components (trapezium type). The inner  0\farcs2 pair A,B (LSC 49) was discovered by \citet{2017AJ....153..212H} in 2012. A fainter star C at 0\farcs5 from A (DSG~13) was first measured in 2016 \citep{2019AJ....157...56H}. The object has been visited four times with Zorro, and C was confirmed at 832\,nm (but never detected in the blue filter) with a magnitude difference of 5.6\,mag (Figure~\ref{fig:57241}). The mean magnitude difference between A and B is 2.98 and 2.70 mag at 562 and 823\,nm, respectively. Observations of this system at SOAR resolved only the pair A,B because component C was below the detection limit. Zorro measured A,C with an angle 180$\degr$ from that of \citet{2017AJ....153..212H}. The position of A,C is stable over time, confirming that this system is bound. The projected $\rho$ imply periods of 50 and 150\,yr for A,B and AB,C, respectively. The pair A,B has turned from  98$\degr$ in 2012 to 40$\degr$ in 2023 with little change in $\rho$, suggesting a quasi-circular face-on orbit with a period of $\sim$70\,yr. The angle of  A,C has changed only by $-$3\fdg5 in 6\,yr, a speed that corresponds to a 600\,yr circular orbit. A plausible dynamically stable configuration would be a wide orbit of C around AB, where C is seen at close $\rho$ owing to projection. In such case its slow motion is natural.

{\bf HIP 59426 = 12114$-$1647 = HD 105913} has been known as a triple with an outer 4\farcs7 pair S~634 (ADS 8444) and an inner spectroscopic subsystem Aa,Ab with a period of 211~days \citep{2019AJ....158..222T}. This triple has been resolved at SOAR and its combined inner orbit was published. Gaia DR3 independently determined an astrometric and SB1 orbit of Aa,Ab, although its amplitude is severely reduced by blending (the mass ratio is 0.87). The purpose of Zorro observations has been to measure the inner subsystem more accurately for mass estimation. Unexpectedly, another faint star Ac has been detected in 2020.04 at 832\,nm at 0\farcs33 with $\Delta I = 5.9$\,mag. The detection is secure in each of the five data cubes. So, Zorro observations reveal this system as a new 3+1 quadruple.

The $\rho$ of Aab,Ac implies a period of 25\,yr. Motion in this orbit should leave a signature in the astrometry and RVs. Indeed, a PM anomaly of 6\,mas\,yr$^{-1}$ has been detected by Brandt \citep{2018ApJS..239...31B, 2019ApJS..241...39B}, and the PMs of A  and B  are substantially different. The 211 day subsystem cannot produce such a large effect, so astrometry indirectly confirms the existence of Ac.  No obvious trend is found in the RVs of the spectroscopic pair measured in 2017--2019. The latest spectrum taken in 2023 also matches the orbit. However, the very first observation in 2008 does differ from the orbit by 3\,km\,s$^{-1}$. The Aab,Ac pair can be presently near elongation, corresponding to a slow RV variation. The RV amplitude can be reduced by the face-on orbit orientation.


\begin{figure}[ht]
\centerline{
\includegraphics[clip, trim=70 330 150 250, width=0.8\textwidth]{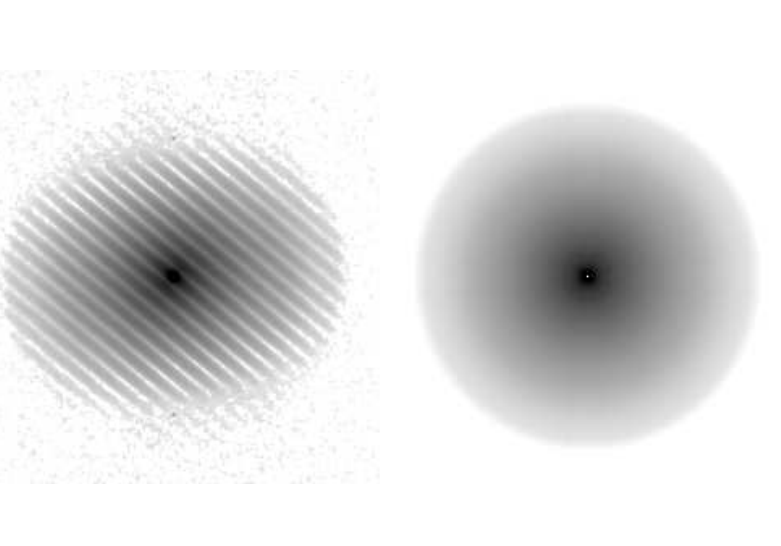} }
\caption{Power spectrum of HIP 58669 (left) and its model (right) taken on 2021.0359 at 832\,nm. The fringes correspond to the A,B pair at a $\rho$ of 0\farcs27. Note the loss of contrast and the inversion of fringe phase in the upper and lower areas, indicating the resolution of Aa,Ab. \label{fig:58669} 
}
\end{figure}

{\bf HIP 58669 = 12018$-$3439  = HD 104471} is a classical solar-type triple system composed of three similar stars. The outer pair I~215 has a combined visual-spectroscopic orbit with $P=156$\,yr, and its main component is a double-lined spectroscopic binary with a period of 148~days \citep{2015AJ....149....8T}. The semimajor axis of Aa,Ab is 13\,mas, and it has been resolved in 2017.43 at similar $\rho$ \citep{2019AJ....157...56H}. We visited this object four times between 2019 and 2022. So far, it has been processed as a binary, but the inner pair is marginally resolved (Figure~\ref{fig:58669}). As we accumulate more epochs, and the data are reprocessed as triple, the measurements of Aa,Ab will allow a determination of its combined orbit, which will define its relative orientation with respect to A,B.  The prospect of accurate mass measurement is however uncertain, considering only marginal resolutions.

{\bf RST 2802 = 12314$-$5659 = HD 108938.} This is a triple-lined system of three similar G8V stars. The outer pair RST~2802 is separated at 1\farcs2 and has an estimated period of 1.4\,kyr. The inner SB2 subsystem Aa,Ab has a period of 343~days, implying a semimajor axis of 8.4\,mas (the parallax of star B is 6.35\,mas, \citet{2022AJ....163..161T}). This system has been observed with Zorro four times in hope of resolving the inner pair. The power spectrum might be slightly elongated, but this resolution is uncertain and we fitted only the binary-star model, measuring the outer 1\farcs25 pair. Our calibration is not accurate enough for detecting astrometric wobble caused by the subsystem with an estimated amplitude of 2\,mas.

{\bf HIP 63162 = 12565$-$2635 = HD 112375} is a visual triple with $\rho$ of 13\farcs8 and 0\farcs15; the inner pair YSC~216 Aa,Ab has been measured here three times. It has turned by $63\degr$ since its discovery in 2011 by \citet{2017AJ....153..212H}. The motion matches the estimated period of 60\,yr, but the orbit is not yet sufficiently constrained to be of any use.

\begin{figure}[ht]
\centerline{
\includegraphics[clip, trim=70 330 150 250, width=0.8\textwidth]{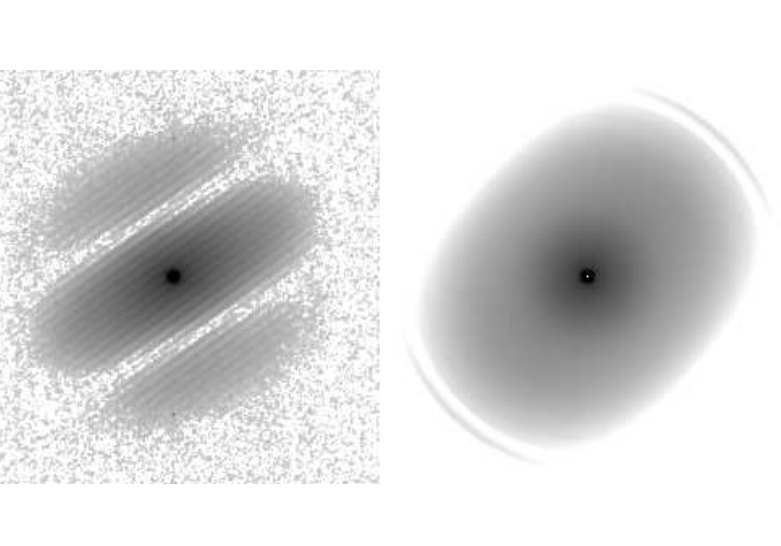} }
\caption{Power spectrum of HIP 63377 taken in 2019.531 at 832\,nm and showing resolution of the inner subsystem. \label{fig:63377} 
}
\end{figure}

{\bf HIP 63377 = 12592$-$6256 =  HD 112636} was noted as double-lined by \citet{2004A&A...418..989N}, and has an astrometric signature (RUWE 10.0 in DR3). The outer 0\farcs45 pair TOK~722 with substantial contrast ($\Delta I = 2.5$\,mag) has been resolved at SOAR in  2016. The inner pair Aa,Ab has been securely resolved by Zorro in 2019, 2021, and 2023 (see Figure~\ref{fig:63377}) and marginally resolved in 2022.37. The estimated period of Aa,Ab is 3\,yr. A spectrum taken with CHIRON in 2023 January clearly shows double lines. This is a good candidate for continued speckle and spectroscopic monitoring for a determination of the inner orbit and masses.

{\bf HIP 68587 = 14025$-$2440 = HD 122445, G3V} is a triple system where the 0\farcs5 visual pair B~263  (period 161\,yr) contains a single-lined spectroscopic subsystem with a period of 2.738\,yr according to Griffin.
who also noted a trend in its center of mass velocity. The corresponding semimajor axis is 31\,mas. However, the full spectroscopic elements remain unpublished, and Roger Griffin passed away. The subsystem Aa,Ab has been tentatively resolved in 2017.42 by \citet{2019AJ....157...56H} at 10\,mas $\rho$ with a small magnitude difference. The object has been visited by Zorro three times and only the A,B  binary was measured. Although the existence of Aa,Ab leaves no doubt, its published resolution remains unconfirmed.

{\bf HIP  72622 = 14509$-$1603  = HD ~130841, A3IV, $\alpha$2~Lib} is a naked-eye quadruple system of 2+2  hierarchy located at 23\,pc. The outer CPM pair has a $\rho$ of 231\arcsec and an estimated period of 170\,kyr. Its secondary component B (HIP 72603,  HD~130819) is a single-lined spectroscopic binary with a period of 14.3\,yr \citep{1991A&A...248..485D} resolved as BEU~19. The primary star A is also a single-lined SB1 with $P=70.3$~days \citep{2014MNRAS.437.2303F} resolved as DSG~17 in 2017 by \citet{2019AJ....157...56H}. It has been resolved in 2019 and 2020 by Zorro at 27 and 18\,mas, respectively, and unresolved in 2023. Using these observations and the published RVs, a visual-spectroscopic orbit of Aa,Ab can be computed. It is unlikely that the orbits of Aa,Ab and Ba,Bb are aligned, given the large outer $\rho$.

\begin{figure}[ht]
\centerline{
\includegraphics[clip, trim=70 330 150 250, width=0.8\textwidth]{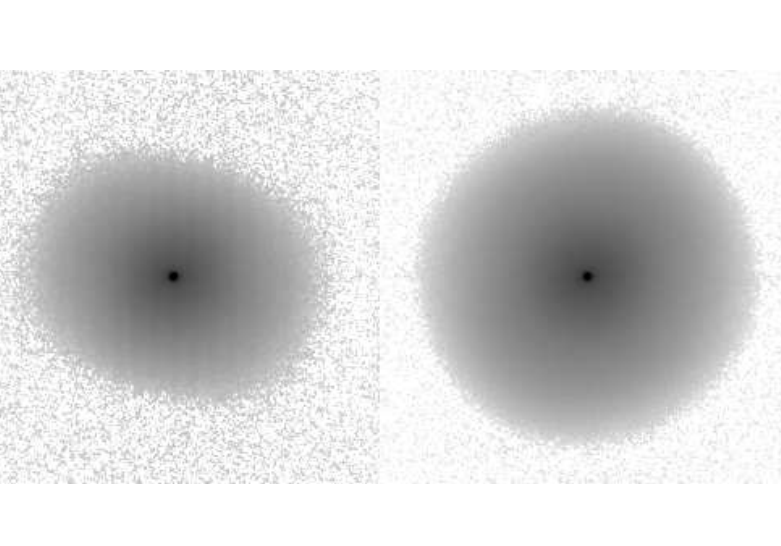} }
\caption{Power spectrum of HIP~76400 taken on 2019.38 at 832\,nm. The fine low-contrast fringes are produced by the companion Ab at 0\farcs21, and the elongation indicates partial resolution of Aa1,Aa2 at 25\,mas. \label{fig:76400} 
}
\end{figure}

{\bf HIP 76400 = 15362$-$0623 = HD 139059} is a 3+1 quadruple system. The outer 80\arcsec pair A,B and the intermediate 0\farcs4 pair Aa,Ab are both designated as TOK~301 in the WDS. Furthermore, star A is a triple-lined spectroscopic system with an inner period of 368.5~days \citep{2022AJ....163..161T}. Its estimated axis of 20\,mas prompted Zorro observations aimed to resolve Aa1,Aa2. In three visits, we securely measured the pair Aa,Ab and noted a slight elongation of the power spectrum due to partially resolved Aa1,Aa2 (Figure~\ref{fig:76400}). Instead of fitting a triple-star model, we estimated the positions of Aa1,Aa2 by fitting a close binary with a $\Delta m =0$ constraint to the red channel only. These positions roughly agree with the spectroscopic elements (see Table~\ref{tab:orbel}).

{\bf HIP 76424 = 15365$+$1607 = HD 139225} is a bright but uninteresting triple with outer $\rho$ of 35\farcs5 (UC~3033) and inner $\rho$ of 1\arcsec (DSG~18 Aa,Ab). The long estimated periods of 50\,kyr and~250\,yr do not favor orbit determination in this century. We measured the inner pair in 2019 and 2020 only in the red channel. Its position has not changed since its discovery in 2014 by \citet{2019AJ....157...56H}. The measured $\Delta m$ at 823\,nm is large (6.4\,mag) 

{\bf HIP~81023 =  16329$+$0315 =  HD~149162} belongs to a late-type quintuple system. The outermost CPM pair A,BC  (LEP~79) has a $\rho$ of 252\arcsec ~and an estimated period of 770\,kyr, while the period of the 6\farcs3 pair B,C (DAM~649) is 7\,kyr. Stars A, B, and C have similar PMs and parallaxes in Gaia DR3. Component A itself hosts three stars: the inner spectroscopic pair Aa,Ab with a period of 226~days \citep{2002AJ....124.1144L} and a low-mass star Ac at 0\farcs3. Both pairs were resolved by Horch in  2013 and bear the name DSG~7 in the WDS. \citet{2022AJ....163..220V} have done a joint spectroscopic (SB1) and astrometric solution, obtaining a semimajor axis of 22\,mas. We observed this target with Zorro in 2019.46, 2019.54, and 2023.17 at zenith distances of 35--39\degr and fitted a binary-star model that corresponds to the intermediate Aab,Ac pair. The power spectrum is slightly elongated, suggesting partial resolution of Aa,Ab, but uncorrected AD prevents reliable estimation of its parameters; data in the blue channel are hopelessly distorted by AD. Given these discouraging results, the object will not be observed regularly in the future. The estimated period of Aab,Ac is 30\,yr, and our measures of this pair will be useful in a future orbit calculation.

{\bf HIP 82730 = 16546$-$0609} is listed in the WDS as a triple system with two companions B and C at $\rho$ of 0\farcs35 and 0\farcs66, respectively, discovered by \citet{2012AJ....144..165H} in 2012.57. However, further measurements between 2015 through 2017 by \citet{2020AJ....159..233H} feature only the A,B pair. This object was observed by Zorro in 2019 twice, and we also do not confirm the existence of additional component C. This system is therefore a simple binary.

{\bf HIP 84430  = 17157$-$0949 = HD 156034} is a resolved triple system where the visual orbits of A,B and Ba,Bb (A~2592 and TOK~53) with periods of 137 and 5.26\,yr, respectively, are well constrained \citep{2021AJ....161..144T}. Our measurement of this triple in 2022.21 and 2023.50 agrees with both orbits.

{\bf HIP 88937 = 18093$-$2607 = HD 165896} is a triple system consisting of the 1\farcs3 visual pair Aa,Ab and a double-lined spectroscopic subsystem  Aa1, Aa2  with a period of 38~days according to its unpublished orbit listed in the MSC. The corresponding semimajor axis of 6\,mas does not favor resolution with Zorro and, indeed, we measured only Aa,Ab in three visits. The estimated period of this pair is 260\,yr, and the observed motion is slow.  WDS lists another two faint companions B and C which are unrelated, as evidenced by the fast relative motion. The sky in this region is very crowded.


{\bf HIP 101472 = 20339$-$2710 =  HD 195719} is a triple system composed of the 52~CPM pair A,B (CBL~178) and the double-lined binary Aa,Ab with a period of 355~days and a mass ratio of 0.92 \citep{2019AJ....158..222T}. Gaia DR3 provided an astrometric orbit of this pair with an inclination of 91\degr (edge-on). The semimajor axis of 14\,mas prompted  Zorro observations aimed to resolve this pair. Observations in 2021.55, 2021.71, and 2022.37 measured $\rho$ of 18, 19, and 15 mas, respectively; the last measurement below the diffraction limit is tentative. An orbit fitted to these data is presented in Figure~\ref{fig:orb}.

\section{Conclusions} \label{sec:concl}

In this paper we report on the astrometric characterization of the Zorro speckle camera mounted on the 8.1\,m GS telescope, based on measurements of binary systems obtained in the context of our survey of southern hemisphere low-metallicity binary and multiple stellar systems. We show that Zorro@GS is a powerful facility for the study of tight visual binary systems and systems with faint secondary components (or the combination of both).

Zorro@GS allows to resolve binary systems even slightly below its natural diffraction limit of~20\,mas, which proved to be critical for the discovery of tight multiple systems among our targets and for the study of the orbital architecture of known compact hierarchical systems we included in our program. 

We show that, with proper calibration, astrometric data ($\rho$, $\theta$ and $\Delta m$) of excellent quality can be obtained. For targets with $\rho$ smaller than 0\farcs4, an overall precision of 1\,mas in the radial and tangential directions is achieved, while the uncertainty in $\theta$ is 0\fdg2. A repeatability study in relative photometry indicates a precision of about 0.1\,mag both in the blue and red filters.

Simultaneous blue (562\,nm) and red (832\,nm) images can be acquired, but the orientation and scale of the two cameras differ, and this has to be considered for high-precision relative astrometry. We compare the $\rho$ and $\theta$ of well-measured binaries with $\rho$ between 50\,mas and 1\farcs2 (typically around 10 such pairs per observing run), which allows us to put them on a common system with an uncertainty smaller than 0.6\%. In this way the blue and red measurements become commensurable, and can be considered equivalent, but independent, in the subsequent orbital analysis. We perform this relative calibration in every observing run, because the instrument sometimes has to be removed from the telescope. In each observing run we also observed at least two astrometric calibration binaries, which allow us to estimate a fractional correction to the absolute plate scale and orientation in the sky. These are long-period (slow motion) bright binaries with $\rho$ between 0\farcs5 and 1\farcs2, that have plenty of modern interferometric measurements, typically from our SOAR program (albeit not necessarily good orbits).



Given certain orbital architectures, however, the limitations of the Zorro@GS setup become evident when the semimajor axis is below the natural diffraction limit of Gemini at 832\,nm, in which case the resolutions do not yield accurate measurements of the positions. Even if more data is secured, the prospect of measuring masses with sufficient (a few percent) accuracy for targets like these is unlikely. We report on two examples of this kind: HIP 76400 (semimajor axis of 19.5\,mas) and HIP~101472 (semimajor axis of 13.9\,mas).

We have also found that, due to the lack of an atmospheric dispersion corrector at Gemini, the blue images are systematically worse than those in the red channel, despite its smaller formal diffraction limit. This means that no detection might be possible in the blue filter if the dispersion happens to be along the $\theta$ of the companion, despite being above the diffraction limit. This is a severe limitation of the instrument in its present form.

Here we present relative astrometry and contrast brightness in both Zorro filters for 70 pairs in 64 distinct systems (six are hierarchical triples). Eleven new binaries have been found (among these, four PSF stars that were considered to be bona-fide single stars until now), most of them with small $\rho$ (down to 15\,mas) and large $\Delta m$ (up to $\Delta m=6$ in the red channel).

We plan to continue observing the most interesting targets with Zorro, as well as adding new promising objects, with the medium-term goal of extending the sample to the low-metallicity regime (currently our lowest metallicity object has [Fe/H]=$-$1.72). In the longer term, we hope that our observations could contribute to the advance of our knowledge of stellar masses for a wide range of metallicities.\newline\newline

RAM, EC and MD acknowledge support from FONDECYT/ANID grant \# 124
0049. RAM also acknowledges support from ANID, Fondo GEMINI, Astrónomo
de Soporte GEMINI-ANID grant \# 3223 AS0002. We acknowledge many
discussions with Dr. Elliot Horch, who provided inspiration, expert
guidance, and interesting targets for GS prior to publication,
especially in the early stages of this work. Finally we thank an
annonymous referee, his/her suggestions and comments were of great
help to improve the readability of our paper.

\vspace{5mm}
\facilities{NOIRLab:Gemini 8.1~m-South+ZORRO, ESO: La Silla, MPG2.2m+FEROS}

\software{IDL}






\bibliography{zorro}{}

\begin{thebibliography}{}
\expandafter\ifx\csname natexlab\endcsname\relax\def\natexlab#1{#1}\fi
\providecommand{\url}[1]{\href{#1}{#1}}
\providecommand{\dodoi}[1]{doi:~\href{http://doi.org/#1}{\nolinkurl{#1}}}
\providecommand{\doeprint}[1]{\href{http://ascl.net/#1}{\nolinkurl{http://ascl.net/#1}}}
\providecommand{\doarXiv}[1]{\href{https://arxiv.org/abs/#1}{\nolinkurl{https://arxiv.org/abs/#1}}}

\bibitem[{{Anguita-Aguero} {et~al.}(2022){Anguita-Aguero}, {Mendez},
  {Claver{\'\i}a}, \& {Costa}}]{2022AJ....163..118A}
{Anguita-Aguero}, J., {Mendez}, R.~A., {Claver{\'\i}a}, R.~M., \& {Costa}, E.
  2022, \aj, 163, 118, \dodoi{10.3847/1538-3881/ac478c}

\bibitem[{{Anguita-Aguero} {et~al.}(2023){Anguita-Aguero}, {Mendez}, {Videla},
  {Costa}, {Vanzi}, {Castro-Morales}, \&
  {Caballero-Valdes}}]{2023AJ....166..172A}
{Anguita-Aguero}, J., {Mendez}, R.~A., {Videla}, M., {et~al.} 2023, \aj, 166,
  172, \dodoi{10.3847/1538-3881/acf297}

\bibitem[{{Benedict} {et~al.}(2016){Benedict}, {Henry}, {Franz}, {McArthur},
  {Wasserman}, {Jao}, {Cargile}, {Dieterich}, {Bradley}, {Nelan}, \&
  {Whipple}}]{2016AJ....152..141B}
{Benedict}, G.~F., {Henry}, T.~J., {Franz}, O.~G., {et~al.} 2016, \aj, 152,
  141, \dodoi{10.3847/0004-6256/152/5/141}

\bibitem[{{Boyajian} {et~al.}(2012{\natexlab{a}}){Boyajian}, {McAlister}, {van
  Belle}, {Gies}, {ten Brummelaar}, {von Braun}, {Farrington}, {Goldfinger},
  {O'Brien}, {Parks}, {Richardson}, {Ridgway}, {Schaefer}, {Sturmann},
  {Sturmann}, {Touhami}, {Turner}, \& {White}}]{2012ApJ...746..101B}
{Boyajian}, T.~S., {McAlister}, H.~A., {van Belle}, G., {et~al.}
  2012{\natexlab{a}}, \apj, 746, 101, \dodoi{10.1088/0004-637X/746/1/101}

\bibitem[{{Boyajian} {et~al.}(2012{\natexlab{b}}){Boyajian}, {von Braun}, {van
  Belle}, {McAlister}, {ten Brummelaar}, {Kane}, {Muirhead}, {Jones}, {White},
  {Schaefer}, {Ciardi}, {Henry}, {L{\'o}pez-Morales}, {Ridgway}, {Gies}, {Jao},
  {Rojas-Ayala}, {Parks}, {Sturmann}, {Sturmann}, {Turner}, {Farrington},
  {Goldfinger}, \& {Berger}}]{2012ApJ...757..112B}
{Boyajian}, T.~S., {von Braun}, K., {van Belle}, G., {et~al.}
  2012{\natexlab{b}}, \apj, 757, 112, \dodoi{10.1088/0004-637X/757/2/112}

\bibitem[{{Brandt}(2018)}]{2018ApJS..239...31B}
{Brandt}, T.~D. 2018, \apjs, 239, 31, \dodoi{10.3847/1538-4365/aaec06}

\bibitem[{{Brandt}(2019)}]{2019ApJS..241...39B}
---. 2019, \apjs, 241, 39, \dodoi{10.3847/1538-4365/ab13b2}

\bibitem[{{Davidson} {et~al.}(2009){Davidson}, {Baptista}, {Horch}, {Franz}, \&
  {van Altena}}]{2009AJ....138.1354D}
{Davidson}, James~W., J., {Baptista}, B.~J., {Horch}, E.~P., {Franz}, O., \&
  {van Altena}, W.~F. 2009, \aj, 138, 1354,
  \dodoi{10.1088/0004-6256/138/5/1354}

\bibitem[{{Duquennoy} \& {Mayor}(1991)}]{1991A&A...248..485D}
{Duquennoy}, A., \& {Mayor}, M. 1991, \aap, 248, 485

\bibitem[{{Eddington}(1924)}]{1924MNRAS..84..308E}
{Eddington}, A.~S. 1924, \mnras, 84, 308, \dodoi{10.1093/mnras/84.5.308}

\bibitem[{{Feiden} \& {Chaboyer}(2012)}]{2012ApJ...757...42F}
{Feiden}, G.~A., \& {Chaboyer}, B. 2012, \apj, 757, 42,
  \dodoi{10.1088/0004-637X/757/1/42}

\bibitem[{{Fuhrmann} {et~al.}(2014){Fuhrmann}, {Chini}, {Barr}, {Buda},
  {Kaderhandt}, {Pozo}, \& {Ramolla}}]{2014MNRAS.437.2303F}
{Fuhrmann}, K., {Chini}, R., {Barr}, A., {et~al.} 2014, \mnras, 437, 2303,
  \dodoi{10.1093/mnras/stt2046}

\bibitem[{{Gafeira} {et~al.}(2012){Gafeira}, {Patacas}, \&
  {Fernandes}}]{2012Ap&SS.341..405G}
{Gafeira}, R., {Patacas}, C., \& {Fernandes}, J. 2012, \apss, 341, 405,
  \dodoi{10.1007/s10509-012-1125-3}

\bibitem[{{Gaia Collaboration}(2022)}]{2022yCat.1357....0G}
{Gaia Collaboration}. 2022, {VizieR Online Data Catalog: Gaia DR3 Part 3.
  Non-single stars (Gaia Collaboration, 2022)}, VizieR On-line Data Catalog:
  I/357. Originally published in: Astron. Astrophys., in prep. (2022)

\bibitem[{{Gratton} {et~al.}(1997){Gratton}, {Fusi Pecci}, {Carretta},
  {Clementini}, {Corsi}, \& {Lattanzi}}]{1997ApJ...491..749G}
{Gratton}, R.~G., {Fusi Pecci}, F., {Carretta}, E., {et~al.} 1997, \apj, 491,
  749, \dodoi{10.1086/304987}

\bibitem[{{Horch} {et~al.}(2012){Horch}, {Howell}, {Everett}, \&
  {Ciardi}}]{2012AJ....144..165H}
{Horch}, E.~P., {Howell}, S.~B., {Everett}, M.~E., \& {Ciardi}, D.~R. 2012,
  \aj, 144, 165, \dodoi{10.1088/0004-6256/144/6/165}

\bibitem[{{Horch} {et~al.}(2009){Horch}, {Veillette}, {Baena Gall{\'e}},
  {Shah}, {O'Rielly}, \& {van Altena}}]{2009AJ....137.5057H}
{Horch}, E.~P., {Veillette}, D.~R., {Baena Gall{\'e}}, R., {et~al.} 2009, \aj,
  137, 5057, \dodoi{10.1088/0004-6256/137/6/5057}

\bibitem[{{Horch} {et~al.}(2015){Horch}, {van Altena}, {Demarque}, {Howell},
  {Everett}, {Ciardi}, {Teske}, {Henry}, \& {Winters}}]{2015AJ....149..151H}
{Horch}, E.~P., {van Altena}, W.~F., {Demarque}, P., {et~al.} 2015, \aj, 149,
  151, \dodoi{10.1088/0004-6256/149/5/151}

\bibitem[{{Horch} {et~al.}(2017){Horch}, {Casetti-Dinescu}, {Camarata},
  {Bidarian}, {van Altena}, {Sherry}, {Everett}, {Howell}, {Ciardi}, {Henry},
  {Nusdeo}, \& {Winters}}]{2017AJ....153..212H}
{Horch}, E.~P., {Casetti-Dinescu}, D.~I., {Camarata}, M.~A., {et~al.} 2017,
  \aj, 153, 212, \dodoi{10.3847/1538-3881/aa6749}

\bibitem[{{Horch} {et~al.}(2019){Horch}, {Tokovinin}, {Weiss}, {L{\"o}bb},
  {Casetti-Dinescu}, {Granucci}, {Hess}, {Everett}, {van Belle}, {Winters},
  {Nusdeo}, {Henry}, {Howell}, {Teske}, {Hirsch}, {Scott}, {Matson}, \&
  {Kane}}]{2019AJ....157...56H}
{Horch}, E.~P., {Tokovinin}, A., {Weiss}, S.~A., {et~al.} 2019, \aj, 157, 56,
  \dodoi{10.3847/1538-3881/aaf87e}

\bibitem[{{Horch} {et~al.}(2020){Horch}, {van Belle}, {Davidson}, {Willmarth},
  {Fekel}, {Muterspaugh}, {Casetti-Dinescu}, {Hahne}, {Granucci}, {Clark},
  {Winters}, {Rupert}, {Weiss}, {Colton}, {Nusdeo}, \&
  {Henry}}]{2020AJ....159..233H}
{Horch}, E.~P., {van Belle}, G.~T., {Davidson}, James~W., J., {et~al.} 2020,
  \aj, 159, 233, \dodoi{10.3847/1538-3881/ab87a6}

\bibitem[{{Iben}(2013)}]{2013sepp.book.....I}
{Iben}, Jr., I. 2013, {Stellar Evolution Physics, Volume 1: Physical Processes
  in Stellar Interiors}

\bibitem[{{Kahler}(1972)}]{1972A&A....20..105K}
{Kahler}, H. 1972, \aap, 20, 105

\bibitem[{{Latham} {et~al.}(2002){Latham}, {Stefanik}, {Torres}, {Davis},
  {Mazeh}, {Carney}, {Laird}, \& {Morse}}]{2002AJ....124.1144L}
{Latham}, D.~W., {Stefanik}, R.~P., {Torres}, G., {et~al.} 2002, \aj, 124,
  1144, \dodoi{10.1086/341384}

\bibitem[{{Mann} {et~al.}(2019){Mann}, {Dupuy}, {Kraus}, {Gaidos}, {Ansdell},
  {Ireland}, {Rizzuto}, {Hung}, {Dittmann}, {Factor}, {Feiden}, {Martinez},
  {Ru{\'\i}z-Rodr{\'\i}guez}, \& {Thao}}]{2019ApJ...871...63M}
{Mann}, A.~W., {Dupuy}, T., {Kraus}, A.~L., {et~al.} 2019, \apj, 871, 63,
  \dodoi{10.3847/1538-4357/aaf3bc}

\bibitem[{{Massey} \& {Meyer}(2001)}]{2001eaa..bookE1882M}
{Massey}, P., \& {Meyer}, M. 2001, in Encyclopedia of Astronomy and
  Astrophysics, ed. P.~{Murdin}, 1882, \dodoi{10.1888/0333750888/1882}

\bibitem[{{Mendez} {et~al.}(2017){Mendez}, {Claveria}, {Orchard}, \&
  {Silva}}]{2017AJ....154..187M}
{Mendez}, R.~A., {Claveria}, R.~M., {Orchard}, M.~E., \& {Silva}, J.~F. 2017,
  \aj, 154, 187, \dodoi{10.3847/1538-3881/aa8d6f}

\bibitem[{{Moe} \& {Di Stefano}(2017)}]{2017ApJS..230...15M}
{Moe}, M., \& {Di Stefano}, R. 2017, \apjs, 230, 15,
  \dodoi{10.3847/1538-4365/aa6fb6}

\bibitem[{{Moe} \& {Kratter}(2018)}]{2018ApJ...854...44M}
{Moe}, M., \& {Kratter}, K.~M. 2018, \apj, 854, 44,
  \dodoi{10.3847/1538-4357/aaa6d2}

\bibitem[{{Nordstr{\"o}m} {et~al.}(2004){Nordstr{\"o}m}, {Mayor}, {Andersen},
  {Holmberg}, {Pont}, {J{\o}rgensen}, {Olsen}, {Udry}, \&
  {Mowlavi}}]{2004A&A...418..989N}
{Nordstr{\"o}m}, B., {Mayor}, M., {Andersen}, J., {et~al.} 2004, \aap, 418,
  989, \dodoi{10.1051/0004-6361:20035959}

\bibitem[{{Reid}(1997)}]{1997AJ....114..161R}
{Reid}, I.~N. 1997, \aj, 114, 161, \dodoi{10.1086/118462}

\bibitem[{{Spada} {et~al.}(2013){Spada}, {Demarque}, {Kim}, \&
  {Sills}}]{2013ApJ...776...87S}
{Spada}, F., {Demarque}, P., {Kim}, Y.~C., \& {Sills}, A. 2013, \apj, 776, 87,
  \dodoi{10.1088/0004-637X/776/2/87}

\bibitem[{{Tighe} {et~al.}(2016){Tighe}, {Tokovinin}, {Schurter},
  {Mart{\'\i}nez}, \& {Cantarutti}}]{2016SPIE.9908E..3BT}
{Tighe}, R., {Tokovinin}, A., {Schurter}, P., {Mart{\'\i}nez}, M., \&
  {Cantarutti}, R. 2016, in Society of Photo-Optical Instrumentation Engineers
  (SPIE) Conference Series, Vol. 9908, Ground-based and Airborne
  Instrumentation for Astronomy VI, ed. C.~J. {Evans}, L.~{Simard}, \&
  H.~{Takami}, 99083B, \dodoi{10.1117/12.2233681}

\bibitem[{{Tokovinin}(2018{\natexlab{a}})}]{2018PASP..130c5002T}
{Tokovinin}, A. 2018{\natexlab{a}}, \pasp, 130, 035002,
  \dodoi{10.1088/1538-3873/aaa7d9}

\bibitem[{{Tokovinin}(2018{\natexlab{b}})}]{2018ApJS..235....6T}
---. 2018{\natexlab{b}}, Astrophysical Journal, Supplement, 235, 6,
  \dodoi{10.3847/1538-4365/aaa1a5}

\bibitem[{{Tokovinin}(2019)}]{2019AJ....158..222T}
---. 2019, \aj, 158, 222, \dodoi{10.3847/1538-3881/ab4c94}

\bibitem[{{Tokovinin}(2021)}]{2021AJ....161..144T}
---. 2021, \aj, 161, 144, \dodoi{10.3847/1538-3881/abda42}

\bibitem[{{Tokovinin}(2022)}]{2022AJ....163..161T}
---. 2022, \aj, 163, 161, \dodoi{10.3847/1538-3881/ac5330}

\bibitem[{{Tokovinin}(2023)}]{2023AJ....165..220T}
---. 2023, \aj, 165, 220, \dodoi{10.3847/1538-3881/acca19}

\bibitem[{{Tokovinin} {et~al.}(2019){Tokovinin}, {Everett}, {Horch}, {Torres},
  \& {Latham}}]{2019AJ....158..167T}
{Tokovinin}, A., {Everett}, M.~E., {Horch}, E.~P., {Torres}, G., \& {Latham},
  D.~W. 2019, \aj, 158, 167, \dodoi{10.3847/1538-3881/ab4137}

\bibitem[{{Tokovinin} \& {Latham}(2020)}]{2020AJ....160..251T}
{Tokovinin}, A., \& {Latham}, D.~W. 2020, \aj, 160, 251,
  \dodoi{10.3847/1538-3881/abbad4}

\bibitem[{{Tokovinin} {et~al.}(2010){Tokovinin}, {Mason}, \&
  {Hartkopf}}]{2010AJ....139..743T}
{Tokovinin}, A., {Mason}, B.~D., \& {Hartkopf}, W.~I. 2010, \aj, 139, 743,
  \dodoi{10.1088/0004-6256/139/2/743}

\bibitem[{{Tokovinin} {et~al.}(2022){Tokovinin}, {Mason}, {Mendez}, \&
  {Costa}}]{2022AJ....164...58T}
{Tokovinin}, A., {Mason}, B.~D., {Mendez}, R.~A., \& {Costa}, E. 2022, \aj,
  164, 58, \dodoi{10.3847/1538-3881/ac78e7}

\bibitem[{{Tokovinin} {et~al.}(2015){Tokovinin}, {Pribulla}, \&
  {Fischer}}]{2015AJ....149....8T}
{Tokovinin}, A., {Pribulla}, T., \& {Fischer}, D. 2015, \aj, 149, 8,
  \dodoi{10.1088/0004-6256/149/1/8}

\bibitem[{{Tokovinin}(1997)}]{1997A&AS..124...75T}
{Tokovinin}, A.~A. 1997, Astronomy and Astrophysics, Supplement, 124, 75,
  \dodoi{10.1051/aas:1997181}

\bibitem[{{Torres} {et~al.}(2010){Torres}, {Andersen}, \&
  {Gim{\'e}nez}}]{2010A&ARv..18...67T}
{Torres}, G., {Andersen}, J., \& {Gim{\'e}nez}, A. 2010, \aapr, 18, 67,
  \dodoi{10.1007/s00159-009-0025-1}

\bibitem[{{Trilling} {et~al.}(2007){Trilling}, {Stansberry}, {Stapelfeldt},
  {Rieke}, {Su}, {Gray}, {Corbally}, {Bryden}, {Chen}, {Boden}, \&
  {Beichman}}]{2007ApJ...658.1289T}
{Trilling}, D.~E., {Stansberry}, J.~A., {Stapelfeldt}, K.~R., {et~al.} 2007,
  \apj, 658, 1289, \dodoi{10.1086/511668}

\bibitem[{{Videla} {et~al.}(2022){Videla}, {Mendez}, {Claver{\'\i}a}, {Silva},
  \& {Orchard}}]{2022AJ....163..220V}
{Videla}, M., {Mendez}, R.~A., {Claver{\'\i}a}, R.~M., {Silva}, J.~F., \&
  {Orchard}, M.~E. 2022, \aj, 163, 220, \dodoi{10.3847/1538-3881/ac5ab4}

\end{thebibliography}
\bibliographystyle{aasjournal}

\end{document}